\documentclass{article}
\usepackage{graphicx}

\begin{document}

\begin{center}
 {\LARGE {\bf The unification of the fundamental interaction within
 Maxwell electromagnetism:}\\
 Model of hydrogen atom.\\
 Gravity as the secondary electric force.\\
 Calculation of the unified inertia force\\
}
\vspace{5 mm}

{\Large \bf L. Neslu\v{s}an}
\indent \\

{\large \em Astronomical Institute, Slovak Academy of Sciences,
 05960 Tatransk\'{a} Lomnica, Slovakia; E-mail: ne@ta3.sk
 {\vspace{5mm}}}


{Presented at the intensive mini-workshop:\\
 GOING BEYOND METRIC: Black Holes, Non-Locality and Cognition;\\
 Tatransk\'{a} Lomnica, Slovakia}\\

{October 3-7, 2010}\\
\end{center}

\newcommand{\gae}{~_{\approx}$\hspace{-2.2mm}$^{>}~}

\large

{\bf Abstract.} Considering two static, electrically charged, elementary
particles, we demonstrate a possible way of proving that all known
fundamental forces in the nature are the manifestations of the single,
unique interaction. We suggest to replace the concept of potential
(intensity of field) with the potential energy (acting force) and
re-define the gauging of integration constants in the Schwarzschild
solution of Einstein field equations. We consider the potential energy
in this context regardless it is gravitational or electric potential
energy. In this way, the electric interaction becomes as generating the
space-time curvature as gravity is assumed to do, in the general
relativity. With the new constants, we sketch how the unique interaction
can be described with the help of an appropriate solution of the
well-known, common Maxwell equations. According the solution, there are
two zones, in the system of two oppositely charged particles, where the
force is oscillating. The first particle can be in a stable, constant
distance from the second particle, between the neighbouring regions of
repulsion and attraction. In an outer oscillation zone, the
corresponding energy levels in the proton-electron systems are identical
(on the level of accuracy of values calculated by the Dirac's equations)
to some experimentally determined levels in the hydrogen atom. Another,
inner oscillation zone will probably explain the quantization of atom
nucleus, since its size is the same as the size of the nucleus. In
addition, the magnitude of the corresponding potential energy rises
typically two orders above the Coulombian behavior in accord with the
,,strongness'' of the strong force, in this region. For each system of
two particles, there is also the zone with the macroscopic, i.e.
monotonous behavior of the force. As well, the solution can be used to
demonstrate that the net force between two assemblies consisting each
(or at least one) of the same numbers of both positively and negatively
charged particles is never zero. A secondary electric force, having the
same orientation as the primary electric force between the oppositely
charged particles, is always present. It can be identified to the
gravity. Finally, the solution of the Maxwell equations can be used to
calculate the inertia force of a particle. The term corresponding to
the first-term electric inertia force is zero, therefore the inertia
force is not proportional to the first-term electric charge. The
consistent formulas for both acting and inertia forces enable to
construct the dimensionless (without gravitational constant, permitivity
of vacuum, etc.) equation of motion.\\


\section{Introduction}

   After the unification of weak and electromagnetic forces, we have
known three fundamental interactions in the nature: gravitational,
electromagnetic, and strong. In the first half of 20-th century,
Einstein suggested the idea of an unification of all interactions. A
radical variant of this idea is not only a formally unified description
of the three forces, but the concept of an actually unique force. A
manifestation of such a force in different circumstances can exhibit
different properties, but the nature and source of the force should be
the same.

   At the present, the phenomena in the macrocosm are described by
the general relativity (Einstein, 1915, 1916) and Maxwell theory of
electromagnetism (Maxwell, 1864). On contrary, the phenomena in the
microcosm are described by the quantum physics. Another unification
attempt is that of the theories of macrocosm and microcosm.

   The attempts of this work is a finding (i) the unique theory for
both macrocosm and microcosm and (ii) the unique force.

   Despite our advanced aim, we present no new principial theory because
it is not necessary for our purpose: the Maxwell theory of
electromagnetism, which was worked out a long time ago, appears to be
absolutely suitable.

   In this work, we suggest some assumptions, changes, and new
representation of few quantities in the classical concepts of current
physics, which could eventually lead to a description of the atom
within the Maxwell theory of electromagnetism and, most probably,
unification of the gravity, strong, and electromagnetic forces. To
demonstrate clearly and in detail an actual success of these
modifications, we do not attempt to create a complete, all comprehending
theory, but we deal, instead, only with the simplest electrostatic
interaction between two point-like, electrically charged particles on
an elementary level of analysis. A full success in the simplest static
configuration of the particles could help in a looking for a further
generalization of the current physical theories.

\section{New representation of Schwarzschild's solution}

   From a certain point of view, an essential difference between the
classical, Newtonian, and general-relativistic concepts of gravity
between two static point-like particles is the existence of non-zero
critical distance, well-known as the Schwarzschild's gravitational
radius. This radius figures in the general relativity, but it is
absent in the Newtonian physics. The relativistic curvature of
space-time, correspodning with gravitational force in classical physics,
diverges when the mutual distance between the particles approaches this
radius. An essential step in our considerations is the suggestion of a
scheme how to introduce an analogue of the Schwarzschild's radius into
a description of electric force.

   In the Schwarzschild solution of the Einstein's field equations,
the components of metric tensor
\begin{equation}\label{g11g44}
g_{44} = -1/g_{11} = C_{1} + \frac{C_{2}}{r},
\end{equation}
where $C_{1}$ and $C_{2}$ are the integration constants. These constants
use to be determined by the demand that
\begin{equation}\label{class2U}
g_{44} = 1 + \frac{2U}{c^{2}},
\end{equation}
in the limit of classical, Newtonian physics. $U = -\kappa m/r$ is the
classical gravitational potential, $c$ is the velocity of light,
$\kappa$ is the gravitational constant, $m$ is the mass of the central
object, and $r$ is the radial distance.

   Below, we define such a way of gauging $C_{2}$ that the factor of 2
in the nominator of fraction $2U/c^{2}$ appears in the case of
macroscopic interaction between two bodies consisting of the same number
of both positively and negatively charged elementary particles. However,
if we consider the interaction between only two such particles, the form
\begin{equation}\label{classU}
g_{44} = 1 + \frac{U}{c^{2}},
\end{equation}
with the single $U$ instead $2U$, is relevant.

   The {\em potential} as well as {\em intensity of field} are
artificially established physical quantities. They can never be
directly measured or observed. In an unification, there occurs the
problem that the physical units, in which the gravitational and
electric potential are expressed, are different (J/kg and J/C,
respectively). We can directly detect only {\em force}, acting between
the objects, and derive the corresponding {\em potential energy.}
Therefore, we replace the potential $U$ with the classical potential
energy, $W_{P}$ (in the classical case, there is always valid
$W_{P} \propto 1/r$), and $c^{2}$ with the rest energy, $W_{o}$. Or,
we replace the ratio $U/c^{2}$ with the ratio $W_{P}/W_{o}$. In our
unified approach to the interaction, we consider potential energy
figuring in the components $g_{11}$ and $g_{44}$ of Schwarzschild
metrique regardless we deal with the electric, gravitational, or other
force.

   The potential energy of a material/charged particle in the force
field of a body consisting of $N_{C}$ particles/charges is the sum
$W_{P1} = \sum_{j=1}^{N_{C}} W_{P1;j}$ and the corresponding ratio
$C_{2}/r$ according our new definition is
$C_{2}/r = W_{P1}/W_{o;1} = \sum_{j=1}^{N_{C}} W_{P1;j}/W_{o;1}$.
If there is a body consisting of $N_{T}$ particles/charges, its
potential energy in the field of the body, which consists of $N_{C}$
particles/charges, is
\begin{equation}\label{WpN}
W_{P} = \sum_{s=1}^{N_{T}} \sum_{j=1}^{N_{C}} W_{Ps;j}.
\end{equation}
The corresponding ratio $C_{2}/r$, in this case, is
\begin{equation}\label{C2/r_2}
\frac{C_{2}}{r} = \sum_{s=1}^{N_{T}}\sum_{j=1}^{N_{C}} \frac{W_{Ps;j}}
 {W_{o;s}}.
\end{equation}

   Constant $C_{2}$ can be identified to the critical distance $R_{S}$,
which we will refer to as the Schwar\-z\-schi\-ld's {\em generalized}
radius. Since the potential energy between two charges of the same
polarity is positive, $R_{S}$ can formally be also negative, in contrast
to the classical gravitational radius, which is only positive like
whatever distance. The above new gauging of constant $C_{2}$ (in the case
of $C_{1}$ we retain the gauging $C_{1} = 1$) is important because it
will be shown, that $R_{S}$ is a limit separating two force regimes,
microscopic and macroscopic.

   The new establishment of the determination of constant $C_{2}$ means
the abandoning of universal field of whatever acting mass/charge, in
fact. In this determination, we must always speak not only about the
mass/charge of the acting object, but we must also characterize the rest
mass and charge of the object particle, for which $C_{2}$ (or $R_{S}$)
is determined.

\section{New features and adopted assumptions}

   In the next section, we outline a concept of hydrogen atom, in which
the electron is in the rest in a fixed distance $r$ from the proton.
This contradicts to the first model of atom in the Bohr's theory that
was the first to explain the atomic energy states. We know, Bohr
postulated the {\em quantum condition} to explain the quantitization of
energetic spectrum of atom. He assumed that the electrons move around
the atomic nucleus in circular orbits, whereby the product of orbital
length, $2\pi r$, and momentum, $m_{oe} v$, is an integer multiple of
the Planck's constant, $h_{B}$, i.e. $2\pi r m_{oe} v = n h_{B}$; $v$ is
the orbital velocity of the electron with the rest mass $m_{oe}$ and $n$
is a positive integer.

   If the true reality is, however, that comprehended in our concept of
static electrons and nucleus, then the factor of $2\pi$ in the Bohr's
condition is redundant. To obtain the correct numerical results in the
experiments with the redundant $2\pi$, the constant $h_{B}$ had also to
be increased about this factor. Therefore, the correct Planck constant,
$h$, equals the original, Bohr-condition-based $h_{B}$ divided with
$2\pi$, i.e. $h = h_{B}/(2\pi )$. Similarly, the Planck's constant
divided by $2\pi$ usually denoted as $\hbar$ must also be corrected:
$\hbar = \hbar_{B}/(2\pi ) = h_{B}/(2\pi )^{2}$.

   It appears that the fine-structure constant has to be corrected by
the factor of $2\pi$ as well. Specifically, $\alpha = 2\pi \alpha_{B}$,
where the $\alpha$ represents the corrected value, while
$\alpha_{B} \doteq 1/137$ is the original value of this constant. The
necessity of the correction is obvious from the definition:
$\alpha = q_{o}^{2}/(4\pi \varepsilon_{o} \hbar c) = 
q_{o}^{2}/\{ 4\pi \varepsilon_{o} [\hbar_{B}/(2\pi )]c\} =
2\pi \alpha_{B}$. In the last formula, symbol $q_{o}$ stands for the
elementary electric charge (charge of proton), and $\varepsilon_{o}$ for
the permitivity of vacuum (SI units are used throughout the text). Our
further suggestions of the correction of $h_{B}$ are given in the
concerning paragraphs.

   The solution of the Maxwell equations, which we find in the sections
below, implies that every particle should be accompanied by a waving
environment. In fact, we accept the quantum-physics concept of a wave
associated with particle. Because of this reason we also accept the
appropriate relations. Specifically, the energy, $W$, of a single
particle in free space can be expressed with the help of the well-known
de Broglie's relation $W = \hbar_{B} \omega$. Here,
$\hbar_{B} = h_{B}/(2\pi )$. Moreover, we assume that the classical
potential energy, $W_{P}$, of the particle should be added to
$\hbar_{B} \omega$ when the particle is situated in a force field, i.e.
\begin{equation}\label{ww1}
W = \hbar_{B} \omega + W_{P}.
\end{equation}

   Similarly, it is accepted that the impulse, $\vec{p}$, is given by
another well-known de Broglie's relation
\begin{equation}\label{htransk}
\vec{p} = \hbar \vec{k},
\end{equation}
where $\vec{k}$ is the wave vector. The magnitude of $\vec{p}$ figures
in the well-known relation for the energy of a moving particle. Adding
also potential energy, we have
\begin{equation}\label{W(p)}
W = \sqrt{p^{2}c^{2} + W_{o}^{2}} + W_{P} =
 \sqrt{k^{2}\hbar^{2}c^{2} + W_{o}^{2}} + W_{P}.
\end{equation}

   The angular frequency $\omega$ has such a behavior that
$\omega \rightarrow \omega_{o}$ for $r \rightarrow \infty$, whereby
$\omega_{o} > 0$. It means that the wave associated with the particle is
an evanescent wave with the amplitude of wave vector, $k$, given by the
well-known relation, which can be written as
\begin{equation}\label{kevan}
k = \frac{\hbar}{\hbar_{B}}\frac{\sqrt{\omega^{2} - \omega_{o}^{2}}}{c} =
 \frac{2\pi}{c} \sqrt{\omega^{2} - \omega_{o}^{2}}
\end{equation}
after the correction of the implicitly present Planck's constant about
the factor of $2\pi$. The angular frequency $\omega_{o}$ of the
associated wave, when the distance between the followed and acting
particles is $r \rightarrow \infty$, can be written in terms of
so-called rest mass of the followed particle, $m_{o}$, with the help of
well-known relation
\begin{equation}\label{m0omega0}
\hbar_{B} \omega_{o} = m_{o} c^{2}.
\end{equation}

   We make a formal unification of the denotation of the electric
charge and mass establishing so-called {\bf electromass}. Having two
particles with charges $q_{T}$ and $q_{A}$, their electromasses $M_{T}$
and $M_{A}$ acquire such the values that Newtonian force between
point-like bodies with masses $M_{T}$ and $M_{A}$ is the same as the
Coulombian force between the particles with charges $q_{T}$ and $q_{A}$,
i.e.
\begin{equation}\label{newtoncoulomb}
\kappa\frac{M_{T}M_{A}}{r^{2}} = \frac{1}{4\pi \varepsilon_{o}}
 \frac{q_{T}q_{A}}{r^{2}},
\end{equation}
or the product of electromasses is
\begin{equation}\label{MTMA}
M_{T}M_{A} = \frac{q_{T}q_{A}}{4\pi \varepsilon_{o}\kappa}.
\end{equation}
In contrast to common mass, the electromass can acquire also negative
values. 

   Special elecromass is {\bf elementary electromass}, $M_{o}$,
corresponding to elementary electric charges, $q_{o}$. It can be given
as
\begin{equation}\label{Mo}
M_{o} = \frac{q_{o}}{\sqrt{4\pi \varepsilon_{o}\kappa}}.
\end{equation}
Numerically, $M_{o} = (1.8594 \pm 0.0003)\times 10^{-9}\,$kg. It is
related to the Planck's mass, $M_{P}$, as
$M_{o} = \sqrt{\alpha_{B}}M_{P}$.

\section{Maxwell's equations for two particles}

   Let us to consider a test particle (TP) and an acting particle (AP),
which acts on the TP. The particles are charged with the electric
charges $q_{T}$ and $q_{A}$, and have rest masses $m_{T}$ and $m_{A}$,
respectively. We consider the static case; the mutual distance between
TP and AP is $r$. Since we consider the static case, it is possible to
assume zero Lorentz magnetic force and not deal with the vector of
magnetic induction, $\vec{B}$.

   The Schwarzschild's generalized radius for the TP in this
two-particle system is
\begin{equation}\label{Re}
R_{S;T} = -\frac{q_{T}q_{A}}{4\pi\varepsilon_{o}m_{T}c^{2}} =
 -\frac{\kappa M_{T}M_{A}}{m_{T}c^{2}}.
\end{equation}
If we consider the hydrogen atom, i.e. TP is an electron and AP is a
proton, then
\begin{equation}\label{Rse}
R_{S;e} = \frac{q_{o}^{2}}{4\pi\varepsilon_{o}m_{e}c^{2}} =
 \frac{\kappa M_{o}^{2}}{m_{e}c^{2}} = 2.8180\times 10^{-15}\, {\rm m}.
\end{equation}

   The electrostatic force between the TP and AP can easily be obtained
from the famous Maxwell equations (MEs) (Maxwell, 1864), when the vector
of the intensity of electric field, $\vec{E}$, is simply multiplied with
the charge $q_{T}$. Keeping this fact in the mind, we use, in the
following, the traditional quantities: intensity of electric field and
electric current, $\vec{J}$, which enable to write the MEs in the
traditional form:
\begin{equation}\label{ME1}
{\rm div} \vec{E} = \frac{\rho}{\varepsilon_{o}},
\end{equation}
\begin{equation}\label{ME2}
{\rm rot} \vec{E} = -\frac{\partial \vec{B}}{\partial t},
\end{equation}
\begin{equation}\label{ME3}
{\rm div} \vec{B} = 0,
\end{equation}
\begin{equation}\label{ME4}
{\rm rot} \vec{B} = \mu_{o}\left( \vec{J} + \varepsilon_{o}
 \frac{\partial \vec{E}}{\partial t}\right) .
\end{equation}

   In free space in the vicinity of TP and AP, we can put the density
of the electric charge $\rho = 0$. The electric current is related to
$\vec{E}$ as
\begin{equation}\label{vecJvecE}
\vec{J} = \zeta \vec{E},
\end{equation}
where $\zeta$ is electric conductivity of vacuum.

   In MEs describing the electromagnetic waves, the non-zero electric
conductivity, $\zeta$, in vacuum is assumed. The relation between the
conductivity, angular frequency, $\omega_{T}$, and magnitude of the wave
vector, $k$, of the TP was found, in the context of the telegraph
equation, as
\begin{equation}\label{zeta}
\zeta = -\frac{i}{\mu_{o}\omega_{T}}\left( k^{2} -
 \frac{\omega_{T}^{2}}{c^{2}}\right) ,
\end{equation}
where $i$ is the unit of imaginary numbers and $\mu_{o}$ is the
permeability of the vacuum.
 
   At the derivation of unique force, we will consider the {\em unit}
intervals of length and time, which correspond to the wave formally
associated with the elementary electromass, $M_{o}$. We define the unit
interval of length identifying this interval to the Compton wavelength
of wave associated with $M_{o}$ divided by the factor of $\pi$, i.e.
distance $L_{o} = \hbar_{B}/(\pi M_{o}c)$. We further define the unit
interval of time, $P_{o}$, as that during which the wave, spreading
with the velocity $c$, overcomes the defined unit distance, i.e.
$P_{o} = L_{o}/c$.

   The corresponding angular frequency of this wave is
\begin{equation}\label{greatomega}
\Omega = \frac{2\pi}{P_{o}}.
\end{equation}
Consequently, we use a modified relation for $\zeta$, in which the
frequency $\omega_{T}$, corresponding to mass $m_{T}$, is replaced with
the frequency $\Omega$ multiplied by the factor of $2\pi$.

   After some usual handing with the MEs, we can convert these equations
to
\begin{equation}\label{mme1}
\Delta \vec{E} - \frac{1}{c^{2}}\frac{\partial^{2} \vec{E}}
 {\partial t^{2}} - \mu_{o}\zeta \frac{\partial \vec{E}}{\partial t}
 = \vec{0}
\end{equation}
for the vector $\vec{E}$, where $\Delta$ is the Laplace's operator.

   In the following, we use the spherical coordinates $r$, $\vartheta$,
and $\varphi$. Considering the common dependence of $\vec{E}$ on time,
but with angular frequency $\Omega$ ($\propto$$M_{o}$) replacing
frequency $\omega_{T}$ ($\propto$$m_{T}$)
\begin{equation}\label{Etime}
\vec{E}(r, \vartheta , \varphi , t) = \vec{E}_{o}(r, \vartheta ,
 \varphi ) \exp(-i 2\pi \Omega t),
\end{equation}
(factor $2\pi$ must be inserted in the argument of the exponential) and
utilizing the introduced modified relation for the conductivity $\zeta$,
\begin{equation}\label{zetamod}
\zeta = -\frac{i}{2\pi \mu_{o}\Omega}\left[ k^{2} -
 \frac{(2\pi \Omega )^{2}}{c^{2}}\right] ,
\end{equation}
we can obtain the wave equation for $\vec{E}$ in the case of the system
of two particles, TP and AP. This vector equation is
\begin{equation}\label{vecEk2}
\Delta \vec{E} + k^{2}\vec{E} = \vec{0}.
\end{equation}

   There can be considered two concepts of manipulation with the
magnitude of the wave vector $k$:\\
(1) $k$ is the explicit function of the radial distance, $r$;\\
(2) $k$ is the implicit function of $r$.\\
It appears that the second alternative implies a plausible solution for
$\vec{E}$ from the point of view of the demands in physics. To
demonstrate a certain link between the presented theory and classical
quantum mechanics, we deal with the first alternative, explicit
$r$-dependence of $k$, at first.

\section{Relationship with quantum mechanics}

   Let us to consider the conditions, in which the classical,
Coulombian, potential energy
$W_{P} = q_{T}q_{A}/(4\pi \varepsilon_{o}r)$ or
$W_{P} = \kappa M_{T}M_{A}/r$ can be applied. Earlier, we introduced the
relation $W = \sqrt{\hbar^{2}c^{2}k^{2} + W_{o}^{2}} + W_{P}$ (relation
(\ref{W(p)})) for the total energy $W$ of TP in the force field of AP.
Using this relation, the quadrate of the wave vector can be given as
\begin{eqnarray}\label{k2}
k^{2} = \frac{W_{o}^{2}}{\hbar^{2}c^{2}}\left[ \left( \frac{W^{2}}
 {W_{o}^{2}} -1\right) + \frac{2WW_{P}}{W_{o}^{2}} + \frac{W_{P}^{2}}
 {W_{o}^{2}}\right] = \nonumber \\
 = \frac{W_{o}^{2}}{\hbar^{2}c^{2}}\left[ \left( \frac{W^{2}}
 {W_{o}^{2}} -1\right) + 2\frac{W}{W_{o}}\frac{\kappa M_{T}M_{A}}
 {W_{o}r} + \frac{\kappa^{2}M_{T}^{2}M_{A}^{2}}{W_{o}^{2}r^{2}}\right] .
\end{eqnarray}
The magnitude of the wave vector, $k$, can also be expressed using the
relation (\ref{W(p)}) in which the energy is put to equal to
$W = W_{o}/\sqrt{g_{44}} + W_{P} = W_{o}/\sqrt{1 + W_{P}/W_{o}} + W_{P}$.
Comparing both relations, one can find
\begin{equation}\label{kuseful}
k = \frac{W_{o}}{\hbar c}  \sqrt{\frac{-\frac{W_{P}}{W_{o}}}
 {1 + \frac{W_{P}}{W_{o}}}}.
\end{equation}

   Let us now to consider only the first ($r$-component) equation of the
vector equation $\Delta \vec{E} + k^{2}\vec{E} = \vec{0}$, i.e. the
equation $\Delta E_{r} + k^{2}E_{r} = 0$ for the radial component
$E_{r}$ of the intensity/force. The explicit form of the equation is
(but with implicit $k^{2}$ for sake of brewity)
\begin{eqnarray}\label{Er1}
\frac{\partial^{2} E_{r}}{\partial r^{2}} + \frac{2}{r}
 \frac{\partial E_{r}}{\partial r} + k^{2}E_{r} + \frac{1}{r^{2}}
 \frac{\partial^{2} E_{r}}{\partial \vartheta^{2}}
 + \frac{\cos \vartheta}{r^{2}\sin \vartheta}\frac{\partial E_{r}}
 {\partial \vartheta} + \frac{1}{r^{2}\sin^{2}\vartheta}
 \frac{\partial^{2} E_{r}}{\partial \varphi^{2}} - \nonumber \\
 - \frac{2}{r^{2}}E_{r} - \frac{2}{r^{2}}\frac{\partial E_{\vartheta}}
 {\partial \vartheta} - \frac{2\cos \vartheta}{r^{2}\sin \vartheta}
 E_{\vartheta} - \frac{2}{r^{2}\sin \vartheta}
 \frac{\partial E_{\varphi}}{\partial \varphi} = 0.
\end{eqnarray}
From this equation for $E_{r}$, the components $E_{\vartheta}$ and
$E_{\varphi}$ can be eliminated using the Gauss' law,
div$\, \vec{E} = \rho /\varepsilon_{o} = 0$. We can find
\begin{eqnarray}\label{eeteep}
-\frac{2}{r^{2}}\frac{\partial E_{\vartheta}}{\partial \vartheta}
 -\frac{2\cos \vartheta}{r^{2}\sin \vartheta}E_{\vartheta} -
 \frac{2}{r^{2}\sin \vartheta} \frac{\partial E_{\varphi}}
 {\partial \varphi} = \frac{2}{r}\frac{\partial E_{r}}{\partial r} +
 \frac{4}{r^{2}} E_{r}.
\end{eqnarray}
After supplying the latter to the wave equation for $E_{r}$, we obtain
\begin{eqnarray}\label{eer2}
\frac{\partial^{2} E_{r}}{\partial r^{2}} + \frac{4}{r}
 \frac{\partial E_{r}}{\partial r} + \frac{2}{r^{2}} E_{r} +
 k^{2} E_{r} + \frac{1}{r^{2}}\frac{\partial^{2}E_{r}}
 {\partial \vartheta^{2}} + \frac{\cos \vartheta}{r^{2}\sin \vartheta}
 \frac{\partial E_{r}}{\partial \vartheta} + \frac{1}
 {r^{2}\sin^{2}\vartheta}\frac{\partial^{2} E_{r}}
 {\partial \varphi^{2}} = 0.
\end{eqnarray}
Now, we can proceed in the same way as people use to proceed in
quantum mechanics: to separate variables and make the Dirac
decomposition of the $r$-coordinate-dependent part of differential
equation of the second degree into two linear differential equations.

   Let us further to apply the wave equation for $E_{r}$ to hydrogen
atom with $M_{T} = -M_{o}$ and $M_{A} = M_{o}$. After the separation of
variables, $E_{r} = \tilde{E}_{r}(r)Y_{r}(\vartheta , \varphi )$, the
$r$-dependent part of the equation (with the already explicitly
expressed quadrate of wave vector, $k^{2}$) is
\begin{equation}\label{KGequ}
\frac{\partial^{2} \tilde{E}_{r}}{\partial r^{2}} + \frac{4}{r}
 \frac{\partial \tilde{E}_{r}}{\partial r} + \frac{2}{r^{2}}
 \tilde{E}_{r} + \left[ \frac{W^{2} - W_{o}^{2}}{\hbar^{2}c^{2}} +
 \frac{2\alpha_{B}W}{\hbar c r} - \frac{l(l + 1) - \alpha_{B}^{2}}
 {r^{2}}\right ] \tilde{E}_{r} = 0,
\end{equation}
where we utilized that $\kappa M_{o}^{2}/(\hbar_{B}c) = \alpha_{B}$.
Number $l$ is integer (requirement yielded from the separation of
variables).

   When we substitute $\psi = \tilde{E}_{r}/r$ in the last equation, it
becomes formally (from the mathematical point of view) identical to the
radial part of the well-known Klein-Gordon equation. (Or, if the
substitution is made before the separation, the presented equation for
$E_{r}$ becomes identical to the original Klein-Gordon equation.) This
identity shows the nature of how the Maxwell electromagnetism and
quantum mechanics are related. If the relation was noticed earlier, the
quantum physics would have, probably, evolved as an extension of the
electromagnetism, not as a new, independent branch of physical
science.\\

\normalsize
   REMARK. The Klein-Gordon equation, which is the basis of Dirac's
equations, is exactly consistent with the formula
$W = \sqrt{p^{2}c^{2} + W_{o}^{2}} + W_{P}$ for the total energy of a
particle. The Dirac's equations would not predict the atomic energy
levels without the correct formula for the energy, which contains the
potential-energy term $W_{P}$. In the classical theory of Schwarzschild
black holes, Oppenheimer \& Volkoff used the Chandrasekhar's equation of
state, which did not include term $W_{P}$ in the description of energy
states of particles of Fermi-Dirac gas. If term $W_{P}$ is included in
the description, one can demonstrate, repeating the Oppenheimer \&
Volkoff (1939) procedure, that the corresponding equation of state (for
the gas-pressure gradient) is
\begin{equation}\label{ES}
\frac{dP}{dr} = -n\frac{dW_{P}}{dr}\left( 1 + \frac{W_{tot}}{W_{o}}
 \sqrt{g_{44}}\right) ,
\end{equation}
where $n$ is the number density, $P$ is the pressure, and $W_{tot}$ is
the total mean energy of gas particles (Neslu\v{s}an, 2009; 2010). The
relativistic condition of thermodynamic equilibrium, i.e. the condition
of stability of a supercritically massive compact object, is
\begin{equation}\label{RETE}
\frac{dP}{dr} = -n\frac{d W_{P}}{dr}
\end{equation}
when expressed with the help of $dP/dr$, $n$, and $dW_{P}/dr$. Equations
(\ref{ES}) and (\ref{RETE}) become identical and object (black hole?) is
stable, when $g_{44} \rightarrow 0$, i.e. when the radius of object
becomes identical to the appropriate gravitational radius.\\

\large
   Since the ME for $E_{r}$ is formally identical to the Klein-Gordon
equation and can further be decomposed into the Dirac's equations, one
solution can be that well-known in the quantum mechanics (ignoring force
and dealing only with energy). Within the Maxwell electromagnetism, we
however consider the common concept of acting electric force. So, we try
to find another solution, which is more consistent with this concept.
We empirically found that if we use the correction of the Planck's
constant larger than or equal to $4\pi$, and expand the components
$E_{r1}$ and $E_{r1}$ of the amplitude of $\tilde{E}_{r}$ (see relation
(\ref{matrixE}) below) into the infinite power series of
$\xi = \alpha_{B} \sqrt{1 - W/W_{o}}\, r/R_{Se}$ (instead of finite
series in classical quantum mechanics), then we obtain the behavior of
the force with some interesting properties (described below). We verify
this as well as all further solutions comparing the calculated and
experimental energy states in the case of the hydrogen atom.

\begin{figure}
\centerline{
   \includegraphics[height=7.5cm,angle=-90]{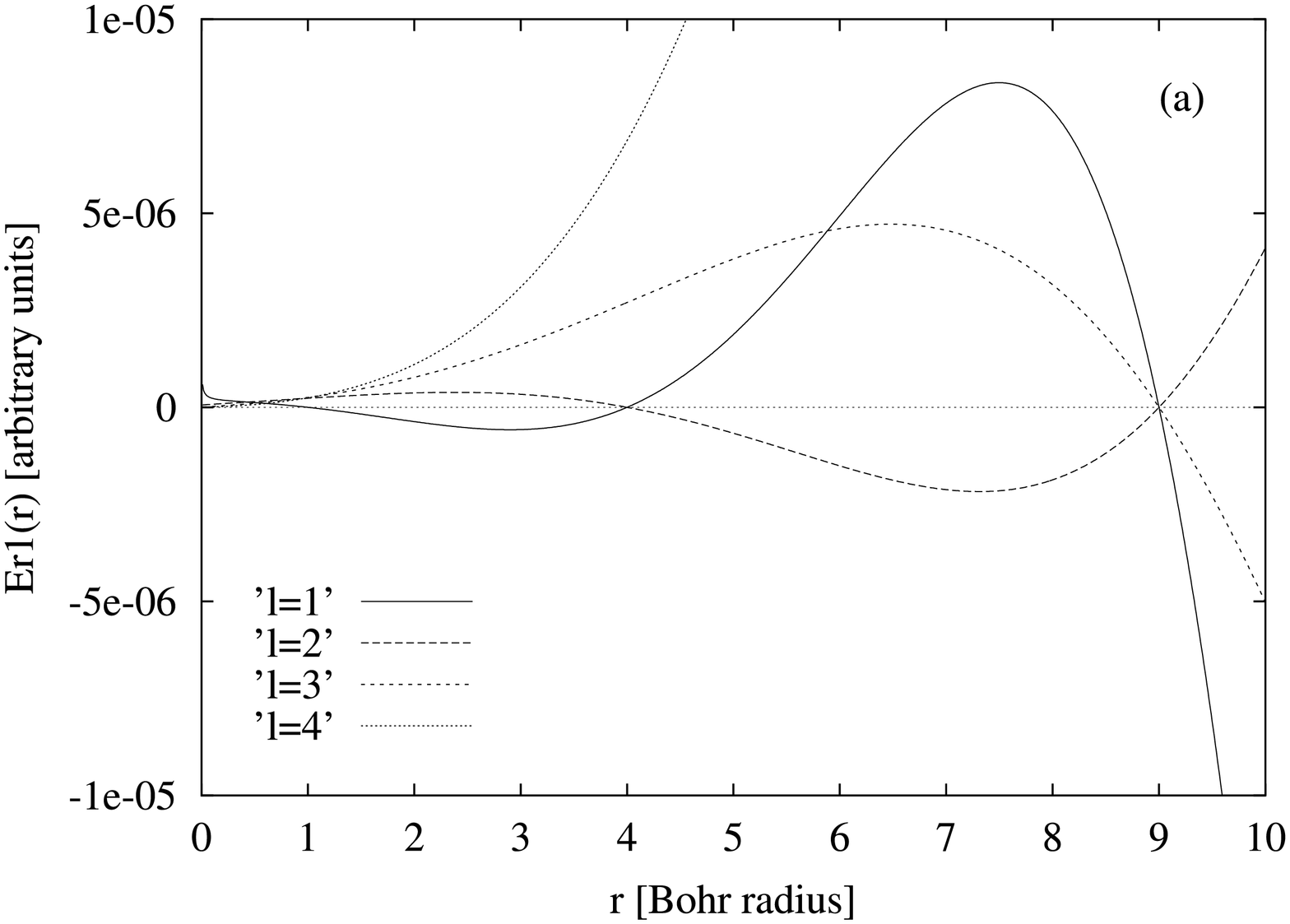}
   \includegraphics[height=7.5cm,angle=-90]{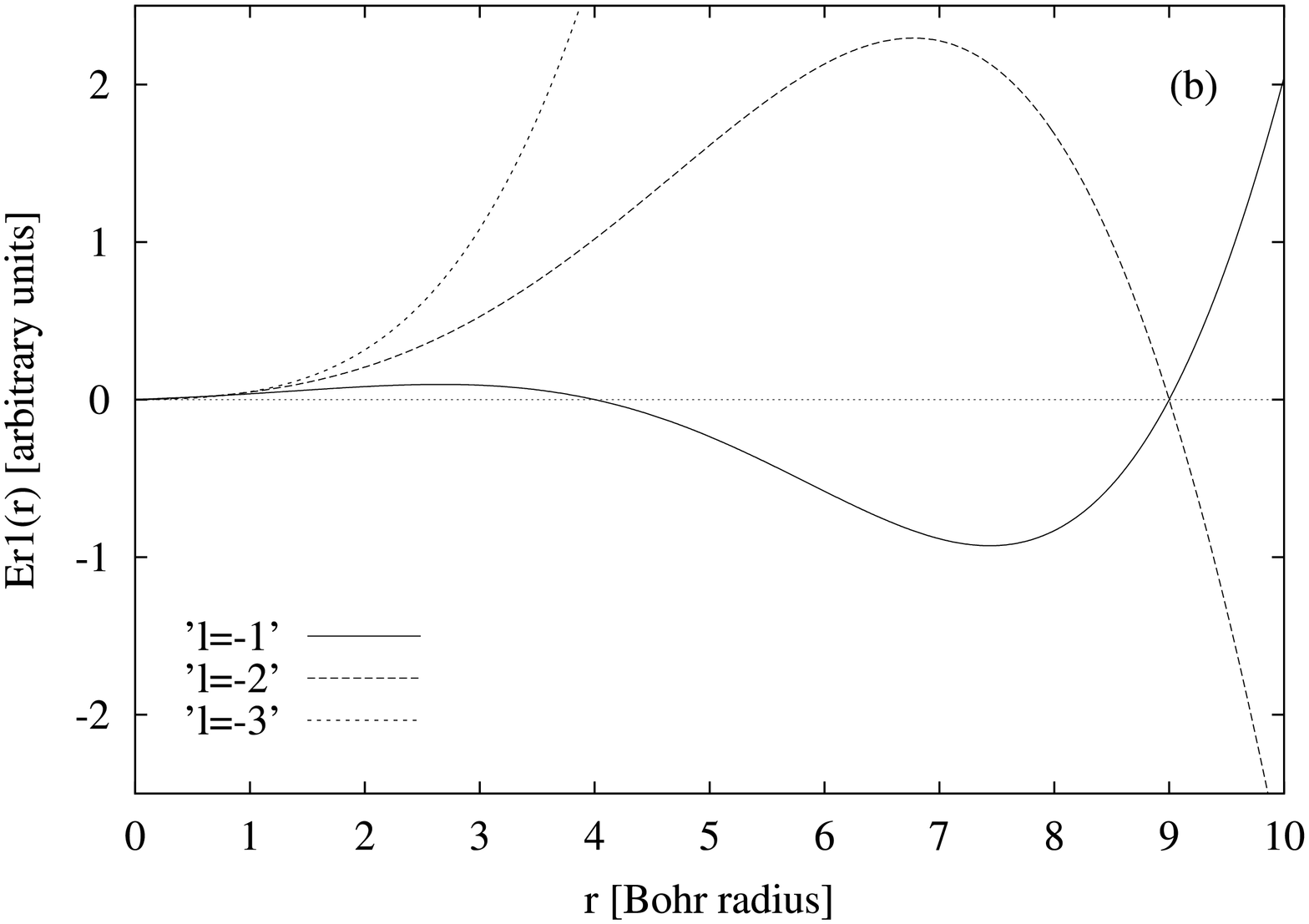}}
\label{fig1}
\caption[f1]{The dependence of the amplitude of intensity of electric
field, $E_{r1}$, in the hydrogen atom on the radial distance, $r$,
from the central proton. The behavior is illustrated for several
first positive (a) as well as negative (b) numbers $l$ in the region
of atomic shell. To demonstrate the amplitude in an acceptable scale,
it is divided by function $\exp (2\xi )$ in the plots.}
\end{figure}

   The behavior of the component $E_{r1}$ in the hydrogen atom is shown
in Fig.~1. For every $l$, the behavior of the force oscillates.
The oscillation implies the zero force at certain, zero-force distances
(ZFDs). For $l = 1$, the ZFDs are approximately equal to 1, 4, 9, 16,
25,... multiple of Bohr's radius ($r_{B}$) (or $r_{0n} \doteq n^{2}r_{B}$
for $n = 1$, 2, 3,...). For $l = 2$ and $l = -1$ (there is no reason why
the negative $l$-numbers should not be considered), ZFDs $\doteq 4$, 9,
16, 25$\, r_{B}$,... For $l = 3$ and $l = -2$, ZFDs $\doteq 9$, 16,
25$\, r_{B}$,... and ZFDs start at higher $n^{2}r_{B}$ for higher $|l|$.

   Summarizing the number of ZFDs through all possible values of $l$, we
can conclude that there is one ZFD at about $1\, r_{B}$, three ZFDs at
$4\, r_{B}$, five ZFDs at $9\, r_{B}$, seven ZFDs at $16\, r_{B}$, etc.
The same ZFDs can be observed in the $E_{r2}$ behavior. So, the correct
number of ZFDs, agreeing with the number of observed discrete energetic
levels in the hydrogen atom, is the result of the solution of the MEs.
No ad hoc quantum condition is necessary to be assumed. We note that
the numerical values of predicted energy levels are practically
identical to the values predicted by the Dirac's theory. A comparison is
given in Table~1.

\begin{table}
\caption[t1]{The ZFDs, $r_{0n}$, and corresponding energetic excesses
over the rest energy, $W - W_{o}$, for several first values of $l$.
The corresponding energetic terms, $W_{D}$, calculated according
the Dirac's theory with the appropriate main quantum number,
$n_{D}$, orbital quantum number, $l_{D}$, and spin-orbital quantum
number, $j_{D}$, as well as the experimental terms \cite{B1},
$W_{\rm exp.}$, are given, too.}
\label{tab1}
\begin{center}
\begin{tabular}{rcc|cccc|c}
\hline \hline
$l$ & $r_{0n}$ & $W - W_{o}$ & $n_{D}$ & $l_{D}$ & $j_{D}$ & $W_{D}$
 & $W_{\rm exp.}$ \\
 & $[r_{B}]$ & $[eV]$ & & & & $[eV]$ & $[eV]$ \\
\hline
 $1$ & 0.9999468 & $-13.598474$ & 1 & 0 & 1/2 & $-13.598474$ & $-13.598439$ \\
 $1$ & 3.9998935 & $-3.3996298$ & 2 & 0 & 1/2 & $-3.3996298$ & $-3.3996253$ \\
$-1$ & 3.9998935 & $-3.3996298$ & 2 & 1 & 1/2 & $-3.3996298$ & $-3.3996297$ \\
 $2$ & 3.9999467 & $-3.3995845$ & 2 & 1 & 3/2 & $-3.3995845$ & $-3.3995843$ \\
 $1$ & 8.9998403 & $-1.5109415$ & 3 & 0 & 1/2 & $-1.5109415$ & $-1.5109402$ \\
$-1$ & 8.9998403 & $-1.5109415$ & 3 & 1 & 1/2 & $-1.5109415$ & $-1.5109414$ \\
 $2$ & 8.9999201 & $-1.5109281$ & 3 & 1 & 3/2 & $-1.5109281$ & $-1.5109280$ \\
$-2$ & 8.9999201 & $-1.5109281$ & 3 & 2 & 3/2 & $-1.5109281$ & $-1.5109280$ \\
 $3$ & 8.9999468 & $-1.5109237$ & 3 & 2 & 5/2 & $-1.5109237$ & $-1.5109236$ \\
 $1$ & 15.999800 & $-0.849902$  & 4 & 0 & 1/2 & $-0.849902$  & $-0.849902$ \\
$-1$ & 15.999790 & $-0.849902$  & 4 & 1 & 1/2 & $-0.849902$  & $-0.849902$ \\
 $2$ & 15.999910 & $-0.849897$  & 4 & 1 & 3/2 & $-0.849897$  & $-0.849897$ \\
$-2$ & 15.999907 & $-0.849897$  & 4 & 2 & 3/2 & $-0.849897$  & $-0.849897$ \\
 $3$ & 15.999939 & $-0.849895$  & 4 & 2 & 5/2 & $-0.849895$  & $-0.849895$ \\
$-3$ & 15.999930 & $-0.849895$  & 4 & 3 & 5/2 & $-0.849895$  & $-0.849895$ \\
 $4$ & 15.999948 & $-0.849894$  & 4 & 3 & 7/2 & $-0.849894$  & $-0.849894$ \\
\hline \hline
\end{tabular}
\end{center}
\end{table}

   The above presented solution for the force between the proton and
electron does not unfortunately satisfy all expectations we demand from
a perfect solution. At first, the amplitude should be a decreasing
function of the increasing radial distance. Instead, the force extremely
increases. For example, the absolute values of local maxima of function
$E_{r1}$ for $l = 1$ between $r_{B}$ and $4r_{B}$, $4r_{B}$ and
$9r_{B}$, $9r_{B}$ and $16r_{B}$ are $1.5\times 10^{-5}$,
$9\times 10^{-3}$, $11$ (in relative units), respectively, for the
correction factor of $4\pi$ (which is also taken ad hoc, and therefore
is not well justified).

   Another shortcoming of the solution found is the fact that the
electron in the atomic shell is in the unstable equilibrium in the half
of ZFDs. Such an unstable ZFD is a transition between the regions of
nearer-to-proton attractive and more-distant-to-proton repulsive forces.
A small external perturbation causes that the electron is moved away
from the ZFD either by the attractive or repulsive force. The electron
is in the stable equilibrium in that ZFD, which is a transition between
the regions of nearer-to-proton repulsive and more-distant-to-proton
attractive force, where the electron is turned back to the ZFD when a
small perturbation occurs. The stable-equilibrium ZFDs in the $E_{r1}$
behavior are unstable-equilibrium ZFDs in the $E_{r2}$ behavior and
vice versa. Combining the $E_{r1}$ and $E_{r2}$ solutions, we can obtain
the complete set of the stable-equilibrium ZFDs necessary to explain all
energy levels, but it is not clear how to swap between $E_{r1}$ and
$E_{r2}$ solutions, when the electron transits from a given
stable-equilibrium ZFD to the neighbouring ZFD.

   At third, we would expect that it should be possible to explain the
macroscopic properties of the electrostatic interaction with the help of
the perfect solution. That means, the amplitude of the intensity should
be Coulombian in some region of distance. However, the obtained
amplitude does not have this property.

\section{Acting force}

\subsection{Acting force generally}

   When looking for the second kind of solution within the Maxwell
electromagnetism, we assume that the magnitude of wave vector, $k$, of
the wave associated with the electron is the {\bf implicit} function of
radial distance, $r$, i.e. $\partial k/\partial r = 0$.

   The equation for the amplitude of the radial component of electric
force, $E_{or}$, is
\begin{equation}\label{eqimpl}
\frac{\partial^{2} E_{or}}{\partial r^{2}} + \frac{4}{r}
 \frac{\partial E_{or}}{\partial r} + \frac{2}{r^{2}} E_{or} +
 k^{2} E_{or} = 0,
\end{equation}
in the case of the proton-electron system. Or, putting
$E_{or} = \tilde{E}_{or}/r^{2}$, we can obtain one-dimensional (scalar)
wave equation in the classical form
\begin{equation}\label{Ek2}
\frac{\partial^{2} \tilde{E}_{or}}{\partial r^{2}} +
 k^{2} \tilde{E}_{or} = 0.
\end{equation}

   The quadrate of wave vector, $k^{2}$, can be decomposed to
\begin{equation}\label{k2kpkm}
k^{2} = k_{+} k_{-},
\end{equation}
where
\begin{equation}\label{kplus}
k_{+} = \frac{2\pi \omega_{o}}{c}\left( \frac{W}{W_{o}} -
 \frac{W_{P}}{W_{o}} + 1\right) ,
\end{equation}
\begin{equation}\label{kminus}
k_{-} = \frac{2\pi \omega_{o}}{c}\left( \frac{W}{W_{o}} -
 \frac{W_{P}}{W_{o}} - 1\right) .
\end{equation}
Using the unit matrix $I_{1}$ and Pauli's matrices $P_{1}$, $P_{2}$, and
$P_{3}$, the wave equation
$\partial^{2} \tilde{E}_{or}/\partial r^{2} + k_{+}k_{-}\tilde{E}_{or} = 0$,
which is the differential equation of the second degree, can be
linearized, i.e. re-written into two linear differential equations
given by
\begin{equation}\label{matrices}
\left(\begin{array}{c} a_{11}~~~ a_{12}\\ a_{21}~~~ a_{22}
 \end{array}\right) \left(\begin{array}{c} \frac{\partial \tilde{E}_{1r}}
 {\partial r}\\ \frac{\partial \tilde{E}_{2r}}{\partial r}\end{array}
 \right) + \left(\begin{array}{c} b_{11}~~~ b_{12}\\ b_{21}~~~ b_{22}
 \end{array}\right) \left(\begin{array}{c} k_{1} \tilde{E}_{1r}\\
 k_{2} \tilde{E}_{2r}\end{array}\right) = \left( \begin{array}{c}
 0\\ 0\end{array}\right) .
\end{equation}
$a_{ij}$ and $b_{ij}$ are the coefficients of an arbitrary combination
of two of matrices $I_{1}$, $P_{1}$, $P_{2}$, and $P_{3}$. Further,
$k_{1} = k_{+}$, $k_{2} = k_{-}$ or $k_{1} = k_{-}$, $k_{2} = k_{+}$
and $\tilde{E}_{1r}$, $\tilde{E}_{2r}$ are the components of matrix
$\tilde{E}_{or}$, i.e.
\begin{equation}\label{matrixE}
\tilde{E}_{or} = \left( \begin{array}{c} \tilde{E}_{1r}\\
 \tilde{E}_{2r}\end{array}\right) .
\end{equation}

   Since there is a lot of combinations of pairs of matrices $I_{1}$,
$P_{1}$, $P_{2}$, and $P_{3}$, moreover doubled with the two possible
combinations of $k_{1}$ and $k_{2}$, one can receive several solutions
of the equation (\ref{eqimpl}). To describe the reality in our universe,
we choose the solution
\begin{equation}\label{stsol}
E_{or} = \frac{K_{E}}{r^{2}}\left[ \pm \sqrt{\frac{k_{+}}{k_{-}}}
 \cos(kr) \mp i\, \sin(kr)\right] ,
\end{equation}
which we refer to, hereinafter, as {\bf the standard solution}. $K_{E}$
is a real-valued integration constant.

   Before we start to analyse the radial force between the TP and AP in
more detail, let us to outline the basic concept of the interaction as
indicated by complete (with the time function) standard solution. Let us
to ask what does the time-dependence part, $\exp(-i 2\pi \Omega t)$,
,,tell'' us. This complex function consists of its real and imaginary
parts:
\begin{equation}\label{timepart}
 \exp(-i 2\pi \Omega t) = \cos(2\pi \Omega t) - i\, \sin(2\pi \Omega t).
\end{equation}
The first knowledge we can gain from the above decomposition is an
indication that we deal with an existence consisting of both real and
imaginary spaces. If we inspect the behavior of the time-dependence part
of radial component of electric-field intensity/force in a point of
space (see relation (\ref{timepart})), we can state the following. In
both real and imaginary spaces, the time-dependent part harmonically
increases and decreases, reaching its positive and negative maximums as
well as zero values. When the real-component (in the real space) reaches
the maximum, the imaginary component (in the imaginary space) is zero
and vice versa.\footnote{In fact, the particle (source of its wave)
oscillating between the real and imaginary spaces can be regarded as
a string. The time-dependence part, $\exp(i\omega t)$, has been assumed
in the solution of the Maxwell equations, as well as the Schr\"{o}dinger
equation in the quantum mechanics, during almost one and half of
century. The concept of strings seems to have a long history in physics,
although the term ,,string'' was not used from the beginning of this
history.}


   The described behavior can be represented by a wave in the real and
wave in the imaginary space (both waves are spherical). The standard
solution for $E_{or}$ contains the wave-vector which has the character
of evanscent wave. This type of wave can transport an energy and carry
an impulse. When the given component of time-dependent part of the above
mentioned behavior is positive, the wave carries an outward oriented
impulse, when it is negative, it carries an inward oriented impulse.
With respect to this, we can guess that the positively-valued wave is
spreading outward (escaping from the center), while the
negatively-valued wave is spreading inward (wave impacting the center).


   Of course, the just described waves must have a source. So, let us to
analyse the behavior of the time-dependent part of the intensity/force
exactly in the place where a considered particle is situated
($r \rightarrow 0$). The behavior is the same as in whatever else point
of space, implying the wave function acquiring both positive and
negative values. This circumstance forces us to assume that the
{\em existence} of the source is as positive as negative (in both real-
as well as imaginary-valued spaces). When the negatively-valued wave
impacts its source being in negative existence, the delivered impulse
has the same direction as the impulse delivered by the positively-valued
wave to the source in the stage of its positive existence. The total,
complex, existence of the particle/source of waving is
$\sqrt{\cos^{2}(2\pi \Omega t) + \sin^{2}(2\pi \Omega t)} = 1$, i.e.
constant.

   In linking all this theory to observations/experiments, it appears
that we are able to observe the phenomena only in the space
corresponding to the real component of our description. The
imaginary-component space seems to be unobservable. So, we will
constrain ourselves to the real-valued parts of found equations or
formulas in every confrontation of our theoretical prediction with
the reality, in the following.

   The fragment of wave escaping and impacting its source, which spreads
in a certain direction, obviously delivers to the source the
corresponding impulse in this direction. If the source/particle is
isolated (no other particle is in its vicinity) and does not accelerate,
the impulse delivered from a given direction is always completely
compensated with the impulse delivered from the opposite direction.
{\bf An interaction occurs when the compensation of the impulse
is broken.} It can happen due to a screening of the associated wave by
an object in the vicinity or due to an acceleration (the impulse from a
red-shifted wave from a given direction is insufficient to compensate
the impulse from a blue-shifted wave from the opposite direction).

   We assume that the size (radius, $R_{I}$) of the source of waves can
be identified to the Compton wavelength, $\Lambda_{C}$, divided by the
factor of $2\pi $ of the wave corresponding to the elementary
electromass: $R_{I} = \Lambda_{C}/(2\pi )$ (or $R_{I} = L_{o}/2$) and,
hence,
\begin{equation}\label{rri}
R_{I} = \frac{\hbar_{B}}{2\pi M_{o}c} = 3.0109\times 10^{-35}\, {\rm m}.
\end{equation}
The interaction takes place just within $R_{I}$, which we also refer to
as the ,,interaction radius''. It is related to so-called Planck's
length, $l_{P} = \sqrt{\kappa \hbar_{B}/c^{3}}$, as
$R_{I} = l_{P}/(2\pi \sqrt{\alpha_{B}})$.

   Since the interaction radius $R_{I} \sim 10^{-35}\,$m is many orders
of magnitude smaller than the smallest generalized Schwarzschild's
radius, i.e. radius between two particles charged with the elementary
electric charges, $R_{S} \sim 10^{-15}\,$m, the standard solution can be
expressed as
\begin{equation}\label{stsolrri}
E_{or} = \frac{K_{E}}{R_{I}^{2}}\left[ \mp i\left(1 + \frac{m_{o}^{2}}
 {2M_{o}^{2}} + ... \right) + \frac{m_{o}}{M_{o}} + ... \right] ,
\end{equation}
in $r = R_{I}$, where we used the equality
$2\pi \omega_{o} R_{I}/c = m_{o}/M_{o}$.

   The amplitude of the force, $M_{o}E_{or}$, multiplied by time part,
$\exp(-i2\pi \Omega t)$, does not provide us with the clear information
about the acting force. The averaging of the force behavior over the
period of time unit is necessary. At the averaging, we use that the
infinitesimal fraction of the amplitude of complex impulse is
$dp = M_{o}E_{or}\, dt$ and the amplitude of the force $F_{or} = dp/dt$.

   Let us now to analyse the situation, when the waves, in both real and
imaginary spaces, associated with the TP are screened by the AP situated
in distance $r$ from the TP. Unfortunately, we must state that the
concept of how the AP absorbes the wave associated with the TP is not
known. To clarify some aspects of unified force, we consider, in the
following, a simple model of the absorption providing at least a rough
quantitative result. The main features of the adopted model are
following. The TP wave is absorbed within the ,,existence sphere'' of
AP, which has the radius $R_{I}$. This sphere is situated in the
distance from $r - R_{I}$ to $r + R_{I}$ from the TP and represents
fraction $(4\pi R_{I}^{3}/3)/(4\pi r^{2}.2R_{I})$ of the volume of TP
wave situated in the above mentioned interval of distance. The
absorption of the TP-wave-carried impulse depends on the measure of AP
existence and the amount of the impulse carried by the TP-wave volume
situated just within the AP sphere of radius $R_{I}$.

   More specifically, we can expect that the measure of the screening
depends on the amplitude of the AP existence which is obviously
proportional to $E_{or}$, related to the AP, at its $R_{I}$ (denotation:
$E_{oR_{I}} = E_{or}(R_{I})$), i.e.
$E_{oR_{I}} = (K_{E}/R_{I}^{2})(\mp i + m_{o}/M_{o})$. We establish
the dimensionless amplitude of the absorption proportional to this
$E_{oR_{I}}$ omitting the constant $K_{E}/R_{I}^{2}$. So, the absorption
amplitude is
\begin{equation}\label{aacp}
A_{AP} = \mp i + m_{o}/M_{o}.
\end{equation}
Finally, the absorbed impulse during the interval $dt$, in our simple,
illustrative model of the absorption, is
\begin{equation}
dp = A_{AP} \frac{4\pi R^{3}/3}{4\pi r^{2}2R_{I}} M_{o}E_{oR_{I}}
 f_{\phi}\, dt,
\end{equation}
where $f_{\phi} = f_{\phi}(t)$ is the function having the meaning
explained below.

  The cosine parts of functions $\exp(-i 2\pi \Omega t)$ and
$\exp(-i 2\pi \Omega t - \phi )$ describe the oscillations of TP wave
and AP itself in the real-valued space. Further, we consider only these
parts. The oscillations of the TP wave may not necessarily be in the
phase with the oscillation of the AP, therefore we consider the phase
shift $\phi$. In the absorption model, we assume that the outward
spreading, positively-valued wave $[\cos(2\pi \Omega t) > 0]$ is
absorbed by the ,,positive'' existence of the AP
$[\cos(2\pi \Omega t - \phi) > 0]$ and negatively-valued, inward
spreading, i.e. on-TP-returning wave $[\cos(2\pi \Omega t) < 0]$ is
absorbed by the ,,negative'' existence of the AP
$[\cos(2\pi \Omega t - \phi) < 0]$. In Fig.~2, the function of the
impulse behavior $\cos(x)$, carried by the TP wave, is shown by the
solid, red curve, and the function of the AP existence absorbing the
impulse, $\cos (x - \phi )$, by the dashed, blue curve. The absorbed
part of the impulse, due to which the force action occurs, is shown by
the violet, hatched area. We here denote $2\pi \Omega t = x$.

\begin{figure}
\centerline{\includegraphics[height=9cm,angle=-90]{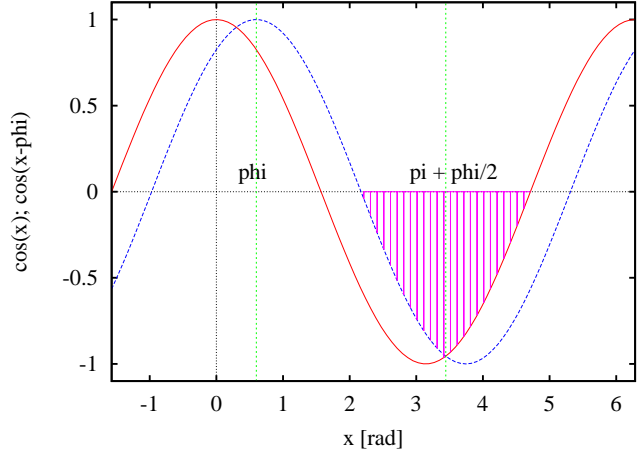}}
\label{fig4}
\caption[f4]{The auxiliary picture to explain the absorption of the
impulse carried by TP wave when the phase shift is $\phi$ (see text
of Sect.~6.1).}
\end{figure}

   The special assumption of our absorption model was already
demonstrated in the last-mentioned figure: if there is more impulse
than AP existence, i.e. $|\cos(x)| > |\cos(x - \phi )|$, then the part
of the impulse equal to the measure of AP existence, $\cos(x - \phi )$,
is absorbed; if the magnitude of AP existence is larger than the actual
magnitude of impulse, i.e. $|\cos(x - \phi )| > |\cos(x)|$, then the
entire actual impulse, $\cos(x)$, is absorbed. Taking this into account,
function $f_{\phi}$, represented by the hatched area in Fig.~2, can be
calculated as
\begin{equation}\label{iiphi}
f_{\phi} = 2 \int_{\pi /2 + \phi}^{\pi + \phi /2} \cos(x - \phi )
 = -2\left( 1 - \sin \frac{\phi}{2}\right) .
\end{equation}
Notice: when $\phi = \pi$, then $f_{\phi} = 0$, i.e no impulse is
absorbed, therefore the AP does not act on the TP.

   In the calculation of integral $f_{\phi}$, we accounted only for the
impulse carried by the returning wave, because only this impulse will
not be delivered to the TP and, thus, will be missing at the
compensation of the impulse from the opposite direction. The size of
the incompensated impulse is proportional to $f_{\phi}$, but is
delivered to the TP from the opposite side than the site of the
absorption of above mentioned returning wave. In the calculation of
the acting force, the impulse proportional to the integral $-f_{\phi}$
is thus relevant.

   Let us now to calculate the impulse absorbed during the whole time
unit interval, $P_{o}$, i.e. the acting force, in fact. This force can
be calculated as the integral over the time of impulse $p$, from 0 to
$P_{o}$, divided by the period $P_{o}$. Since the proper time integral
was already given as the integral $-f_{\phi}$, the acting force can be
expressed as
\begin{equation}\label{ffact1}
F_{act.} = I_{\phi} A_{AP} \frac{4\pi R_{I}^{3}/3}{4\pi r^{2}2R_{I}}
 M_{o} E_{oR_{I}}.
\end{equation}
We denoted $-f_{\phi}/P_{o} = I_{\phi}$.

   In the statistics of a large number of particles, the phase shift
$\phi$ varies randomly from 0 to $2\pi$. Within our simple model of
the absorption, the mean value $<$$I_{\phi}$$>$ of integral $I_{\phi}$
is
\begin{equation}\label{meaniiphi}
<I_{\phi}> = \frac{1}{2\pi}\int_{0}^{2\pi} \frac{1}{\pi} \left( 1 -
 \sin \frac{\phi}{2}\right) \, d\phi = \frac{1}{\pi}\left( 1 -
 \frac{2}{\pi}\right) .
\end{equation}

\subsection{Acting force between macroscopic objects}

   We already gave the amplitude of the intensity, $E_{or}$, (relation
(\ref{stsolrri})) for $r \ll R_{S}$. In the microcosm, there is no
practical use of this formula, since no phenomena are observed in any
scale that is many orders of magnitude lower than $\sim$$10^{-15}\,$m.
In the microcosm, the amplitude of $E_{or}$ is oscillating, proportional
to harmonic functions $\cos(kr)$ and $\sin(kr)$. The concept of the
oscillating force seems to be in a disagreement with the behavior of
the electric force observed in the macroscopic experiments, where the
force is monotonous, either attractive or repulsive, i.e. its sign does
not change. This disagreement can, however, be easily explained when we
realize that the region of distances $r > |R_{S}|$ is attributed to the
atom (i.e. microcosm) and region $r < |R_{S}|$ to the macroscopic
phenomena. Namely, the harmonic functions in the formulas for the
amplitude, $\cos(kr)$ and $\sin(kr)$, change to the monotonous functions
$\cosh(|k|r)$ and $\sinh(|k|r)$ for $r < |R_{S}|$ (implying
$|W_{P}/W_{o}| > 1$). This is clear from the following analysis of the
formula for the magnitude of wave vector (relation (\ref{kuseful})). For
the particles of the same polarity, $W_{P} > 0$, therefore
\begin{eqnarray}\label{kequpolar}
k = \frac{W_{o}}{\hbar c}\sqrt{\frac{-\frac{W_{P}}{W_{o}}}
 {1 + \frac{W_{P}}{W_{o}}}} = i\, \frac{W_{o}}{\hbar c}
 \sqrt{\frac{\frac{W_{P}}{W_{o}}}{1 + \frac{W_{P}}{W_{o}}}} = i |k|.
\end{eqnarray}
And, for the particles of opposite polarity with $W_{P} < 0$, we can
demonstrate that
\begin{eqnarray}\label{koppopolar}
k = \frac{W_{o}}{\hbar c}\sqrt{\frac{-\frac{W_{P}}{W_{o}}}
 {1 + \frac{W_{P}}{W_{o}}}} = \frac{W_{o}}{\hbar c}
 \sqrt{\frac{\left| \frac{W_{P}}{W_{o}}\right| }{1 - \left| 
 \frac{W_{P}}{W_{o}}\right| }} = \frac{W_{o}}{\hbar c}\frac{1}{i}
 \sqrt{\frac{\left| \frac{W_{P}}{W_{o}}\right| }{\left| \frac{W_{P}}
 {W_{o}}\right| -1 }} = -i |k|.
\end{eqnarray}
We now clearly see that $\cos(\pm i|k|r) = \cosh(|k|r)$ and
$\sin(\pm i|k|r) = \pm i\, \sinh(|k|r)$.

   Because of a larger simplicity and, thus, transparency, we will deal
with the Newtoni\-an/Coulombian region of the interaction, further. For
$r \ll |R_{S}|$, hence $|W_{P}/W_{o}| \gg 1$,
$|k| = W_{o}/(\hbar c) = 2\pi m_{o}c/\hbar_{B} = 2\pi \omega_{o}/c$
(with a relatively high precision). The explicit form of magnitude of
the amplitude $E_{or}$ for $r \ll |R_{S}|$ is
\begin{equation}\label{Coulombee1}
|E_{Ar}| = \frac{K_{E}}{r^{2}}\left| \mp i + \frac{2\pi \omega_{o}}{c}
 r + ... \right| \doteq \frac{K_{E}}{r^{2}}.
\end{equation}
It means that the amplitude is not only monotonous, but it becomes
Newtonian/Co\-ulo\-mbi\-an.

   At a first glance, it may seem to be strange that the region of
distances $r > |R_{S}|$ is regarded as a domain of microscopic phenomena
and that with $r \ll |R_{S}|$ as a domain of macroscopic phenomena in
classical physics. This apparent paradox can easily be explained, when
we realize the consequences of the newly established determination of
the generalized Schwarzschild radius. Namely, this radius is
$R_{Se} \sim 10^{-15}\,$m for, e.g., the hydrogen atom. However for
some macroscopic objects, for example those having the charge
$\pm 10^{-9}\,$C, we can found that $R_{S} \sim 10^{5}\,$m. This
distance can actually be regarded as much larger than the common
laboratory distances of decimetres to metres scale. Term
{\em macroscopic} thus primarily means a {\em macroscopic number}
(of charge carriers).

   So, let us now to deal with a test object (TO), which consists of
a large number of positively charged particles, $N_{+}$, of the same
mass $m_{+}$ and large number of negatively charged particles, $N_{-}$,
with the same mass $m_{-}$. In a distance $r$ satisfying $r \ll R_{S}$
($R_{S}$ of the system), let there is situated an acting object (AO)
again consisting of a large number of positively charged particles,
$n_{+}$, having mass $m_{+}$ and large number of negatively charged
particles, $n_{-}$, with mass $m_{-}$. We consider the point-like TO
and AO, i.e. their dimensions are much smaller than their mutual
distance $r$.

   Obviously, the acting force is the appropriate multiple of partial
forces acting on the individual particles of the TO, i.e.
\begin{eqnarray}\label{ffmacro}
F_{act.} = \frac{M_{o}K_{E}<I_{\phi}>}{6r^{2}} \left[ i\, n_{+}
 \left( 1 + \frac{m_{+}^{2}}{2M_{o}^{2}}\right) - i\, n_{-}
 \left( 1 + \frac{m_{-}^{2}}{2M_{0}^{2}}\right) + n_{+}\frac{m_{+}}
 {M_{o}} + n_{-}\frac{m_{-}}{M_{o}}\right] . \nonumber \\
 . \left[ i\, N_{+}
 \left( 1 + \frac{m_{+}^{2}}{2M_{o}^{2}}\right) - i\, N_{-}
 \left( 1 + \frac{m_{-}^{2}}{2M_{0}^{2}}\right) + N_{+}\frac{m_{+}}
 {M_{o}} + N_{-}\frac{m_{-}}{M_{o}}\right] 
\end{eqnarray}
where we already generalized the absorption amplitude, which is also
the appropriate multiple of partial absorptions. The last force law can
be re-written to form (further, we neglect the terms containing the
second power $m^{2}/M_{o}^{2}$)
\begin{eqnarray}\label{ffmacro2}
F_{act.} = \frac{M_{o}K_{E}<I_{\phi}>}{6r^{2}} \left[ i\, (n_{+} -
 n_{-}) + \frac{n_{+}m_{+} + n_{-}m_{-}}{M_{o}} \right] . \nonumber\\
 . \left[ i\, (N_{+} - N_{-}) + \frac{N_{+}m_{+} + N_{-}m_{-}}{M_{o}}
 \right] = \nonumber \\
 = \frac{M_{o}K_{E}<I_{\phi}>}{6r^{2}} \left\{ \left[
 -(n_{+} - n_{-})(N_{+} - N_{-}) + \right. \right. \nonumber\\
 \left. \left. + \frac{1}{M_{o}^{2}} (n_{+}m_{+} +
 n_{-}m_{-})(N_{+}m_{+} + N_{-}m_{-}) \right] + \right. \nonumber \\
 \left. + i\, \frac{1}{M_{o}} \left[ (N_{+} - N_{-})(n_{+}m_{+} +
 n_{-}m_{-}) + (n_{+} - n_{-})(N_{+}m_{+} + N_{-}m_{-}) \right]
 \right\} .
\end{eqnarray}

   In the real world, we assume that only the real component of this
force is observed. Therefore, the macroscopic unified acting force can
be, finally, written as
\begin{eqnarray}\label{ffa}
F_{A} = \frac{M_{o}K_{E}<I_{\phi}>}{6r^{2}} \left[ -(n_{+} - n_{-})
 (N_{+} - N_{-}) + \right. \nonumber \\
 \left. + \frac{1}{M_{o}^{2}}(n_{+}m_{+} + n_{-}m_{-})
 (N_{+}m_{+} + N_{-}m_{-})\right] .
\end{eqnarray}
We note that, usually, $m_{+} \equiv m_{p}$ and $m_{-} \equiv m_{e}$,
therefore ratio $m_{\pm} /M_{o} \ll 1$. Despite this circumstance, we
keep the term containing the quadrate of this ratio, because the first
term may be zero, when $n_{+} = n_{-}$ or $N_{+} = N_{-}$.

   The analysis of force $F_{A}$ yields the following conclusions.\\
(1) Two objects charged with the charges of the same polarity, i.e.
$n_{+} > n_{-}$ and $N_{+} > N_{-}$ (or $n_{+} < n_{-}$ and
$N_{+} < N_{-}$): the first term, which we starts to refer to as ,,the
first electric-force term'', is $-(n_{+} - n_{-})(N_{+} - N_{-}) < 0$.
On contrary, the term $(n_{+}m_{+} + n_{-}m_{-})(N_{+}m_{+} + N_{-}m_{-})
/M_{o}^{2}$, which we will refer to as ,,the secondary electric-force
term'', is positive. Therefore the orientation of the secondary electric
force is opposite with respect to the orientation of the first-term,
dominant electric force. We note that the secondary electric force is
$\sim$$m^{2}/M_{o}^{2} \sim 10^{-36}$ times smaller, when
$m_{+} \equiv m_{p}$ and $m_{-} \equiv m_{e}$.\\
(2) Two objects charged with the charges of the opposite polarity, i.e.
$n_{+} > n_{-}$ and $N_{+} < N_{-}$ (or $n_{+} < n_{-}$ and
$N_{+} > N_{-}$): the first electric-force term, $-(n_{+} - n_{-})
(N_{+} - N_{-})$ is larger than zero, this time. The secondary
electric-force term, $(n_{+}m_{+} + n_{-}m_{-})(N_{+}m_{+} + N_{-}m_{-})
/M_{o}^{2}$, remains positive. Therefore the orientation of the
secondary electric force is the same as the orientation of the
first-term electric force.\\
(3) The charged TO and electrically neutral AO (or vice versa), i.e.
$n_{+} \neq n_{-}$ and $N_{+} = N_{-}$ (or $n_{+} = n_{-}$ and
$N_{+} \neq N_{-}$): the first electric-force term, $-(n_{+} - n_{-})
(N_{+} - N_{-})$ is zero. However, the secondary electric-force term,
$(n_{+}m_{+} + n_{-}m_{-})(N_{+}m_{+} + N_{-}m_{-})/M_{o}^{2}$, remains
the same size and positive as in the previous cases. There is no
first-term electric force acting, but the secondary electric force is
not influenced with the electric charge of whatever of the two
objects.\\
(4) Two electrically neutral objects, i.e. $n_{+} = n_{-}$ and
$N_{+} = N_{-}$: the first electric-force term, $-(n_{+} - n_{-})
(N_{+} - N_{-})$ is zero as expected. However, the secondary
electric-force term, $(n_{+}m_{+} + n_{-}m_{-})(N_{+}m_{+} + N_{-}m_{-})
/M_{o}^{2}$, remains the same size and positive even in this case.
Again, there is no first-term electric force acting, but the secondary
electric force is, as we could see, always ,,surviving''.

   We can see that the nominator of the secondary electric-force term,
$(n_{+}m_{+} + n_{-}m_{-})(N_{+}m_{+} + N_{-}m_{-})$, is, in fact, the
product of multiplication of the masses, $n_{+}m_{+} + n_{-}m_{-}$ and
$N_{+}m_{+} + N_{-}m_{-}$, of the AO and TO, respectively. While the
primary electric force is proportional to unity (constant) and cannot
depend on the curvature of space-time or velocity of the object (in the
sense as the mass depends on the curvature and/or velocity according
$m = m_{o}/\sqrt{g_{44}}$), the secondary electric force does so.

{\bf The revealed secondary electric force has all attributes
to be identified to the gravitational force.}

   We formally assign the cosine (hyperbolic cosine) part of the
electric-intensity amplitude to the electric part (electric charge) of
the unified interaction and sine (hyperbolic sine) part to the
gravitational part (mass) of the interaction.

\subsection{Net electric charge of stars}

   There is a case when the second power of ratio $m/M_{o}$ is not
negligible in the unified force law (in the parentheses of relation
(\ref{ffmacro})). Let us to consider the TP consisting of a single
electron and a large macroscopic body (as the AO) consisting of the
same number, $n_{+}$, of protons and electrons, $n_{-}$ ($n_{+} = n_{-}$),
e.g. a pure-hydrogen star. The absorption amplitude is
\begin{eqnarray}\label{absorpsun}
A_{AO} = n_{+}\, i\, \left( 1 + \frac{m_{p}^{2}}{2M_{o}^{2}}\right)
 + n_{-}(-i)\left( 1 + \frac{m_{e}^{2}}{2M_{o}^{2}}\right) + \nonumber \\
 + n_{+} \frac{m_{p}}{M_{o}} + n_{-}\frac{m_{e}}{M_{o}} =
 i\, n_{+}\frac{m_{p}^{2} - m_{e}^{2}}{2M_{o}^{2}} + n_{+}
 \frac{m_{p} + m_{e}}{M_{o}}.
\end{eqnarray}
The acting force in this system is
\begin{equation}\label{actesun}
F_{act.} = \frac{M_{o}K_{E}<I_{\phi}>}{6r^{2}} A_{AO} \left( -i +
 \frac{m_{e}}{M_{o}}\right) 
\end{equation}
and its component in the real space, the effect of which is observable,
can be obtained, after the product $A_{AO}$ and parentheses is
calculated, as
\begin{equation}\label{actesun2}
F_{A,real} = \frac{M_{o}K_{E}<I_{\phi}>}{6r^{2}} \left( n_{+}
 \frac{m_{p}^{2} - m_{e}^{2}}{2M_{o}^{2}} + n_{+} \frac{m_{p} +
 m_{e}}{M_{o}}\frac{m_{e}}{M_{o}}\right) .
\end{equation}
The third electric-force term (it is written as the first term in the
parentheses) is, here, greater than the secondary (gravitational) term.

   Specifically, the ratio is
\begin{equation}\label{esunratio}
\frac{n_{+} \frac{m_{p}^{2} - m_{e}^{2}}{2M_{o}^{2}}}
 {n_{+} (m_{p} + m_{e}) \frac{m_{e}}{M_{o}^{2}}} =
 \frac{m_{p} - m_{e}}{2m_{e}} \doteq \frac{1}{2}\frac{m_{p}}{m_{e}}
 \doteq 918.
\end{equation}
The third electric-force term is the term of the electric part
(hyperbolic cosine) of the unified force. The above derived force is
the electric force between the net charge of the macroscopic body and
the charge of electron.

   The net electrostatic field of the Sun and other stars was found
within the old model of solar atmosphere by Dutch astronomer Pannekoek
(1922) and generalized as the attribute of the whole solar/stellar body,
not only of atmosphere, by Rosseland (1924). The field was studied till
the middle of 19-ty fifties by several astrophysicists (e.g. Eddington,
1926; Cowling, 1929; Pikel'ner, 1948; 1950; van de Hulst, 1950; 1953).
It was mainly linked, in the minds of experts, to the models of the
solar corona. It seems that it was later forgotten together with the
obsolete models of the corona, when these models were replaced by the
MHD models. The principle of the generation of this field is, however,
not exclusively related to any stellar atmosphere. The lighter and thus
faster moving electrons tend to separate from the heavier protons (ions)
in the stellar plasma. If the Sun was electrically neutral, $22\%$ of
all electrons on its surface would reach the escape velocity (in
contrast to only $10^{-1735}\%$ of protons). When the field acts on an
electron, the electric force is actually $m_{p}/(2m_{e}) \doteq 918$
times stronger than the gravity, as we found.

   The Pannekoek-Rosseland electric field equalizes the probability of
the escape of both polarity charge-carriers away from a star. The field
can be expressed in form of the net charge of star (Neslu\v{s}an, 2001).
This charge is positive and within the sphere of radius $r$ equals
\begin{equation}\label{qqsun}
Q_{r} = \frac{2\pi \varepsilon_{o}\kappa (m_{p} - m_{e})}{q_{o}} M_{r},
\end{equation}
where $M_{r}$ is the stellar mass within the radius $r$. When the third
term of the acting unified force is re-written with the help of product
of charges of electron and macroscopic body, we obtain the charge of
the body identical to the Pannekoek-Rosseland charge (\ref{qqsun}).

   In the representation when the magnitudes of positive and negative
elementary electric charges are exactly the same, the net charge of star
can occur due to the asymmetry of the numbers of carriers of both kinds.
In our concept of unified force and, consequently, the unified concept
of charge and mass, the elementary charge slightly depends on the mass
of its carrier, therefore the magnitude of the charge of proton is
slightly larger than that of electron. In this concept, the
Pannekoek-Rosseland net charge occurs at exactly the same number of
carriers of both positive and negative charges. In the new concept, the
symmetry of the magnitudes of positive and negative elementary charge is
replaced with the symmetry of their numbers.

\subsection{Acting force in microcosm}

   Let us again to consider the proton-electron system. To give the
force acting on the electron in this system, we use the originally
derived relation for the electric intensity (\ref{stsol}), so called
standard solution,
$E_{or} \propto \sqrt{k_{+}/k_{-}}\cos(kr) - i\, \sin(kr)$.
The corresponding force is
\begin{equation}\label{ffmicro}
F_{A,micro} = A_{AP}\frac{M_{o}K_{E}}{6r^{2}}\left[
 \sqrt{\frac{k_{+}}{k_{-}}}\cos (kr) - i\, \sin (kr)\right] ,
\end{equation}
where the absorption amplitude $A_{AP}$ is the same as in the case of
macroscopic interaction (relation (\ref{aacp})) since the proton
absorbes the impulse carried by the wave associated with the electron
with the maximum screening radius equal to $R_{I}$. With this amplitude,
i.e. with $A_{AP} = i + m_{p}/M_{o}$, we can derive
\begin{eqnarray}\label{ffmicro2}
F_{A,micro} = \frac{M_{o} K_{E}}{6r^{2}} \left\{ \left[
 \sin (kr) + \frac{m_{p}}{M_{o}}\sqrt{\frac{k_{+}}{k_{-}}}\cos (kr)
 \right] + \right. \nonumber \\
 \left. + i\, \left[ \sqrt{\frac{k_{+}}{k_{-}}}\cos (kr) -
 \frac{m_{p}}{M_{o}}\sin (kr) \right] \right\} .
\end{eqnarray}
Its real componet, which is manifested in the observed world, is
\begin{equation}\label{ffmicro3}
F_{A,micro} = \frac{M_{o} K_{E}}{6r^{2}} \left[ \sin (kr) +
 \frac{m_{p}}{M_{o}}\sqrt{\frac{k_{+}}{k_{-}}}\cos (kr)\right] .
\end{equation}

   In the atom shell, where $r \approx R_{S}/\alpha_{B}^{2}$,
it is valid $(m_{p}/M_{o})\sqrt{k_{+}/k_{-}} \approx
m_{p}/(\alpha_{B}M_{o}) \ll 1$. So, the second term can well be
neglected, therefore the force becomes proportional only to
the function of $\sin (kr)$, i.e.
\begin{equation}\label{ffmicro4}
F_{A,real} = \frac{M_{o} K_{E}}{6r^{2}} \sin (kr).
\end{equation}
The behavior of $F_{A,real}$ (zero-force distances) is illustrated
in Fig.~3.

\begin{figure}
\centerline{
   \includegraphics[height=9 cm,angle=-90]{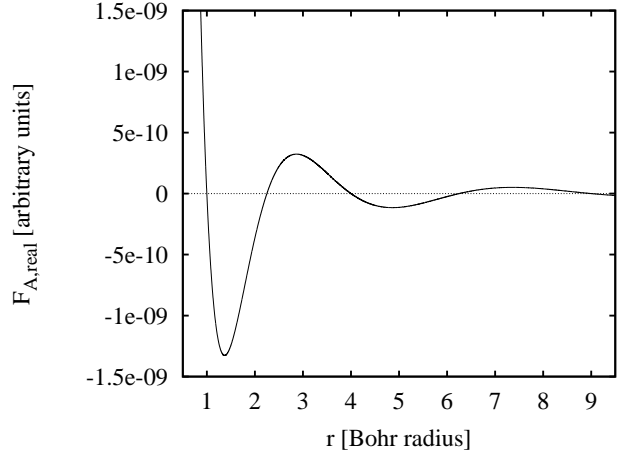}} 
\label{fig5}
\caption[f5]{The dependence of the real-valued amplitude of radial
component of the electric force, $F_{A,real}$, in the hydrogen atom on
distance $r$ from the central proton. The behavior is illustrated for
$l = 0$ in the region of the atomic shell (scale
$r_{B} = R_{Se}/\alpha_{B}^{2}$). The repulsive force corresponds with
$F_{A,real} > 0$ and attractive force with $F_{A,real} < 0$.}
\end{figure}

   From relation (\ref{ffmicro4}), it is clear that the
stable-equilibrium ZFDs, $r_{0n}$, satisfy condition
$k r_{0n} = 2\pi n$, where $n$ is a positive integer. In fact, number
$n$ is an analogue of the main quantum number in the
classical-quantum-mechanics description of hydrogen atom.

   The magnitude of the wave vector, $k$, is given by relation
(\ref{kuseful}) in the general case, or can be written as
\begin{equation}\label{k_class}
k = \frac{2\pi \alpha_{B}}{R_{Se}}\sqrt{\frac{\frac{R_{Se}}{r}}
 {1 - \frac{R_{Se}}{r}}},
\end{equation}
for the proton-electron system. Supplying $k$ into the condition
$k r_{0n} = 2\pi n$ for the ZFDs, we can re-write this condition to
the quadratic equation
\begin{equation}\label{atomequ}
\frac{\alpha_{B}^{2}}{n^{2}}\left( \frac{r_{0n}}{R_{Se}}\right) ^{2}
 - \frac{r_{0n}}{R_{Se}} + 1 = 0
\end{equation}
having the solution 
\begin{equation}\label{r12n}
r_{0n} = \left( 1 \pm \sqrt{1 - 4\alpha_{B}^{2}/n^{2}}\right)
 \frac{n^{2}}{2\alpha_{B}^{2}}R_{Se}.
\end{equation}
The first group of $r_{0n}$, with plus sign in front of the square root
in (\ref{r12n}), corresponds to the stable-equilibrium ZFDs of the
electron in the atomic shell.

   Using relation $W = \sqrt{\hbar^{2}c^{2}k^{2} + W_{o}^{2}} + W_{P}$
(identical to $W = W_{o}/\sqrt{g_{44}} + W_{P}$) for the energy, $W$, in
which we supply $W_{P} = -W_{o}R_{Se}/r$, energy $W$ can be re-written
as
\begin{equation}\label{ww2}
W = W_{o}\left( \frac{1}{\sqrt{1 - R_{Se}/r_{12}}} - \frac{R_{Se}}
 {r_{12}}\right) ,
\end{equation}
where $r_{12} = (1 + m_{oe}/m_{op})r$ includes the correction for the
barycenter of the system ($m_{oe}$ and $m_{op}$ are the rest masses of
electron and proton, respectively). The energy terms can be calculated
supplying $r = r_{0n}$. The obtained enegry terms are identical to those
characterized with $l_{D} = n_{D} - 1$ and $j_{D} = l_{D} + 1/2$ in the
Dirac's solution. (Unfortunately, the full set of energy levels is not
known within the presented implicit-$k$ solution of MEs. It can likely
be obtained after the generalization of the universal metrique towards
the axial-symmetry, e.g. Kerr solution of Einstein field equations and
after the full solution of the MEs, with also $E_{\vartheta}$ and
$E_{\varphi}$ components of the electric-intesity vector found.) The
comparison of our result to the terms predicted according the Dirac's
theory as well as to the experimental values is given in Table~2.

\begin{table}
\caption[t2]{The ZFDs, $r_{0n}$, and corresponding energetic excesses
over the rest energy, $W - W_{o}$, for several first values of $n$
as calculated by the solution for $F_{A,real.}$. The corresponding
energetic terms, $W_{D}$, calculated according the Dirac's theory with
the appropriate main quantum number, $n_{D}$, orbital quantum number,
$l_{D}$, and spin-orbital quantum number, $j_{D}$, as well as the
experimental terms, $W_{\rm exp.}$, are given, too.}
\label{tab1}
\begin{center}
\begin{tabular}{rcc|cccc|c}
\hline \hline
$n$ & $r_{0n}$ & $W - W_{o}$ & $n_{D}$ & $l_{D}$ & $j_{D}$ & $W_{D}$
 & $W_{\rm exp.}$ \\
 & $[r_{B}]$ & $[eV]$ & & & & $[eV]$ & $[eV]$ \\
\hline
 $1$ & 0.9999468 & $-13.598474$ & 1 & 0 & 1/2 & $-13.598474$ & $-13.598439$ \\
 $2$ & 3.9999467 & $-3.3995845$ & 2 & 1 & 3/2 & $-3.3995845$ & $-3.3995843$ \\
 $3$ & 8.9999468 & $-1.5109237$ & 3 & 2 & 5/2 & $-1.5109237$ & $-1.5109236$ \\
 $4$ & 15.999948 & $-0.849894$  & 4 & 3 & 7/2 & $-0.849894$  & $-0.849894$ \\
\hline \hline
\end{tabular}
\end{center}
\end{table}

   When we consider the sign minus in the relation (\ref{r12n}) for
$r_{0n}$, we gain another, second group of the ZFDs. In this case,
$r_{0n}$ can be approximated as
\begin{equation}\label{nuclZFDs}
r_{0n} = R_{Se}\left( 1 + \frac{\alpha_{B}^{2}}{n^{2}} + ...\right)
 \doteq R_{Se}.
\end{equation}
We clearly see that these ZDFs are closely above the generalized
Schwarzschild's radius, at distances $\approx 10^{-15}\,$m. These ZFDs
provide us with a possibility to explain, on the basis of electric
interaction, the quantization in the atomic nucleus. The behavior of
the oscillating force at the Schwarzschild's radius is shown in Fig.~4.

   The possible identity of the strong force with the electric one can
also be supported by the following estimate of energy. In a much larger
or a much smaller distance $r$ than $R_{S}$ (in the case of atom, in
the atom shell), the magnitude of free energy,
$|W - W_{o}| = |W_{o}/\sqrt{|1 - R_{S}/r|} - W_{o}R_{S}/r|$, is of
order $|W - W_{o}| \sim W_{o}R_{S}/r$, i.e. its behavior is
Newtonian/Coulombian. We remind that the energy of radiation (of an
emitted photon) can originate, in atom, from the difference $W - W_{o}$.
At the nucleus, $r \rightarrow R_{Se}$ and we can show that the
magnitude of free energy is of order $\sim$$W_{o}/\alpha_{B}$. Since
the corresponding Newtonian/Coulombian free energy would be of order
$\sim$$W_{o}R_{Se}/R_{Se} \sim W_{o}$, it is clear that the actual free
energy, $W - W_{o}$, is about the factor of $1/\alpha_{B} \sim 10^{2}$
higher at the nucleus in comparison with Coulombian behavior.

\begin{figure}
\centerline{
   \includegraphics[height=9 cm,angle=-90]{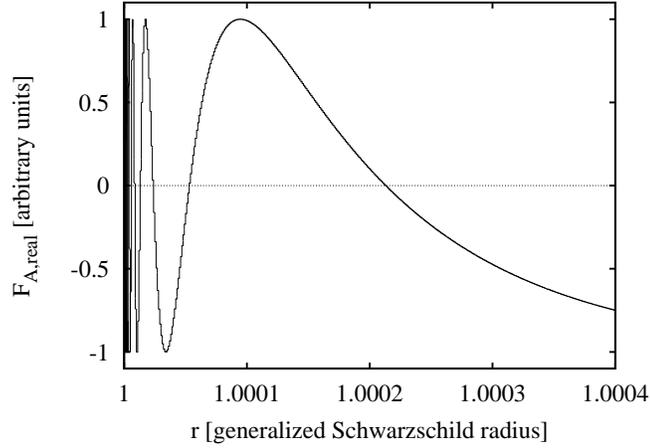}} 
\label{fig6}
\caption[f6]{The dependence of the real-valued amplitude of radial
component of the electric force, $F_{A,real}$, in the hydrogen atom
on distance $r$ from the central proton. In this plot, the behavior is
illustrated in the region close above the {\em generalized Schwarzschild
radius}, $R_{Se}$. Force is given in arbitrary units which, however,
correspond to those used in Fig.~3.}
\end{figure}

   It seems that the old concept of the neutron as the particle composed
of proton and electron becomes again actual. In principle, the energy
excess between the rest energy of neutron and sum of proton+electron
mass could be predicted, here. As well, the nuclear systems consisting
of several protons and neutrons (neutrons as protons+electrons) seems
possible to be explained. Unfortunately, an usage of our concept for
quantitative calculations appears to be difficult, because it is not
longer valid that $r \gg R_{Se}$ (as in atomic shell), therefore the
particles cannot be regarded as point-like objects and, especially,
the dependence of the potential energy on the distance $r$ is not known.
The common Coulomb approximation is absolutely inappropriate in this
ultra-relativistic region.

   At last, we can explain why the ,,strong force'' (electrical, in
fact) does not oscillate and, thus, does not create the stable systems
between two protons (why there is no proton-proton bound pair). The
potential energy of a proton in the electric force field of another
proton is positive, therefore the amplitude of the wave vector, in this
case, is
\begin{equation}\label{k_pp}
k = \frac{2\pi \alpha_{B}}{R_{Sp}}\sqrt{\frac{-\frac{R_{Sp}}{r}}
 {1 + \frac{R_{Sp}}{r}}} = i\, \frac{2\pi \alpha_{B}}{R_{Sp}}
 \sqrt{\frac{\frac{R_{Sp}}{r}}{1 + \frac{R_{Sp}}{r}}} = i|k|,
\end{equation}
where $R_{Sp}$ is the generalized Schwarzschild radius for the
proton-proton system. Therefore, the argument of the functions of
cosine and sine, $kr$, is changed to $i|k|r$. Consequently, the
amplitude of the standard solution now is
\begin{equation}\label{stsolpp}
E_{A,pp} = \frac{K_{E}}{r^{2}}\left[ \sqrt{\frac{k_{+}}{k_{-}}}
 \cosh (|k| r) + \sinh (|k|r)\right] .
\end{equation}
Both hyperbolic cosine and hyperbolic sine are monotonous functions,
which yield no ZFD.

   Finding the stable-equilibrium ZFDs, we create, in fact, the concept
of the atom, in which the electrons in its shell and, probably, the
nucleons in its nucleus are not orbiting the center of the system, but
all these particles are in the rest in the ZFDs. An essential feature of
the concept is the electric force changing its magnitude and sign with
the radial distance from the center of the system, i.e. the oscillating
force.

\section{Inertia force}

   If a particle is in the rest or moves with a constant velocity, its
associated wave is perfectly spherically symmetric and the particle is
exactly in the wave center. The impulse, carried by this wave, is
completely compensated when the wave leaves or impacts the particle,
 as already mentioned.


   However, if the particle accelerates (see scheme in Fig.~5), the wave
in the direction of the acceleration is blue-shifted, its frequency
increases and, consequently, the impulse becomes larger in this
direction. In the opposite direction, the wave is red-shifted and
corresponding impulse reduced. The impulse coming from the direction of
the acceleration of particle is not longer completely compensated by
the impulse from the opposite direction $-$ there occurs
a force acting on the particle. This mechanism is
{\bf the nature of the inertia force.}

\begin{figure}
\centerline{
  \includegraphics[height=3cm,angle=0]{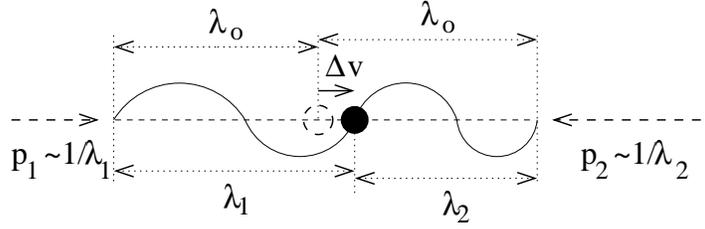}} 
\label{fig8}
\caption[f8]{The impulse carried by the wave associated with the
particle (full bullet) when the particles accelerates (1D case).
This time, the impulse is not completely compensated
($p_{1} < p_{2}$ because $\lambda_{1} > \lambda_{2}$).}  
\end{figure}

   The angular frequency of the wave spreading in the direction declined
by the angle $\vartheta$ from the direction of the acceleration is
\begin{equation}\label{omegavartheta}
\omega_{o} = \frac{\omega_{oc}}{1 - (\Delta v/c)\cos \vartheta},
\end{equation}
where $\omega_{oc}$ is the angular frequency of the particle in rest and
in free space and $\Delta v$ is the change of the velocity of particle
during a time interval. We identify this interval to the unit interval of
time, $P_{o}$.

   The impulse carried by the wave in the direction characterized by
common spherical angles $\varphi$ and $\vartheta$ at the distance of
interaction radius, $R_{I}$, which crosses the infinitesimal area
$R_{I}^{2}\sin \vartheta \, d\varphi \, d\vartheta $ within the time
interval $dt$, according the standard solution, is
\begin{eqnarray}\label{impphitheta}
d^{3}p(\varphi , \vartheta ) = \frac{R_{I}^{2}\sin \vartheta \,
 d\varphi\, d\vartheta}{4\pi R_{I}^{2}} \frac{M_{o}K_{E}}{R_{I}^{2}}
 \left[ \mp i\, \cosh \left( \frac{2\pi \omega_{o}R_{I}}{c}\right) +
 \right. \nonumber \\
 \left. + \sinh \left( \frac{2\pi \omega_{o}R_{I}}{c}\right) \right] 
 \cos(2\pi \Omega t)\, dt.
\end{eqnarray}
In the last relation, we consider only the cosine part of the function
$\exp(-i 2\pi \Omega t)$ describing the wave in the real-valued space
(just this part was also considered at the derivation of acting force).

   The component of the impulse delivered to the particle in the
direction opposite to its acceleration obviously is $d^{3}p(\varphi ) =
\cos \vartheta \, d^{3}p(\varphi , \vartheta )$ (see Fig.~6). The sum
of the impulse over all values of $\varphi$, i.e. from the circle
inclined by angle $\vartheta$ to the aceleration direction, is
$d^{2}p = 2\pi d^{3}p(\varphi , \vartheta )$.

\begin{figure}
\centerline{
  \includegraphics[height=5cm,width=5.75cm,angle=0]{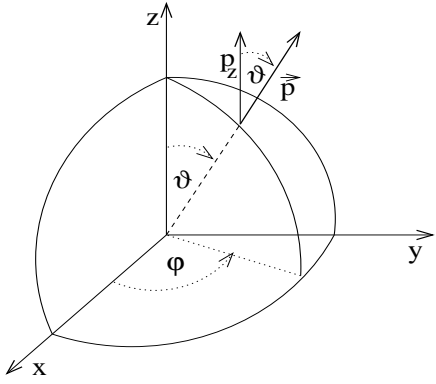}}
\label{fig9}
\caption[f9]{The impulse carried by the wave associated with the
particle: its fraction $\vec{p}$ in the direction characterized by
the angles $\varphi$ and $\vartheta$ and the component of this impulse
$p_{z}$ in the direction of the particle acceleration.}
\end{figure}

   The inertia force is double integral of impulse $d^{2}p$ over
all values of angle $\vartheta$ and over the whole period of the
particle oscillation divided by this period, i.e.
\begin{eqnarray}\label{ffi1}
F_{iner.} = -\frac{1}{P_{o}}\int_{0}^{P_{o}} \int_{0}^{\pi}
 \frac{2\pi R_{I}^{2}}{4\pi R_{I}^{2}}\frac{M_{o}K_{E}}{R_{I}^{2}}
 \left[ \mp i\, \cosh \left( \frac{2\pi R_{I}}{c}\frac{\omega_{oc}}
 {1 - (\Delta v/c)\cos \vartheta }\right) + \right. \nonumber \\
 \left. + \sinh \left( \frac{2\pi R_{I}}{c}\frac{\omega_{oc}}
 {1 - (\Delta v/c)\cos \vartheta }\right) \right] 
 \cos \vartheta \sin \vartheta \, d\vartheta \, 
 | \cos(2\pi \Omega t) |\, dt.
\end{eqnarray}
We add the sign minus to respect the fact that the impulse delivered to
the considered particle has the opposite direction than the acceleration
of the particle.

   If the wave impacts the particle, it gives the carried impulse in
the direction of its motion; if the wave leaves the particle, the
impulse is again given in this direction due to the action-reaction
principle. This fact is respected taking the absolute value
$|\cos(2\pi \Omega t)|$ in the calculation of $F_{iner.}$. If we denote
\begin{equation}\label{iit}
I_{t} = \frac{1}{P_{o}}\int_{0}^{P_{o}}| \cos(2\pi \Omega t)|\, dt =
 \frac{2}{\pi},
\end{equation}
then the force can be given as
\begin{eqnarray}\label{ffi2}
F_{iner.} = -\frac{M_{o}K_{E}I_{t}}{2R_{I}^{2}} \int_{0}^{\pi}
 \left[ \mp i\, \cosh \left( \frac{2\pi R_{I}}{c}\frac{\omega_{oc}}
 {1 - (\Delta v/c)\cos \vartheta }\right) + \right. \nonumber \\
 \left. + \sinh \left( \frac{2\pi R_{I}}{c}\frac{\omega_{oc}}
 {1 - (\Delta v/c)\cos \vartheta }\right) \right] 
 \cos \vartheta \sin \vartheta \, d\vartheta .
\end{eqnarray}
Integral
\begin{eqnarray}\label{iivartheta1}
I_{\vartheta} = \int_{0}^{\pi} \left[ \mp i\, \cosh \left(
 \frac{2\pi R_{I}}{c}\frac{\omega_{oc}}{1 - (\Delta v/c)
 \cos \vartheta }\right) + \right. \nonumber \\
 \left. + \sinh \left( \frac{2\pi R_{I}}{c}\frac{\omega_{oc}}
 {1 - (\Delta v/c)\cos \vartheta }\right) \right] \cos \vartheta
 \sin \vartheta \, d\vartheta 
\end{eqnarray}
can be calculated after we develop the functions hyperbolic cosine
and hyperbolic sine into the power series.
The result of integration is
\begin{eqnarray}\label{iithetafull}
I_{\vartheta} = \pm i\, \left[ S^{2}\sum_{j=1}^{\infty}\frac{2j}
 {2j+1}\left( \frac{\Delta v}{c}\right) ^{2j-1} + \right. \nonumber \\
 \left. + 4\sum_{s=1}^{\infty}\frac{S^{2s}}{(2s)! (2s-1)!}
 \sum_{j=1}^{\infty}\frac{j(2s+2j-2)!}{(2j+1)!}\left(
 \frac{\Delta v}{c}\right)^{2j-1}\right] + \nonumber \\
 + \left[ 2S \sum_{j=1}^{\infty} \frac{1}{2j+1}\left( \frac{\Delta v}
 {c}\right) ^{2j-1} + \right. \nonumber \\
 \left. + 4\sum_{s=1}^{\infty} \frac{S^{2s-1}}{(2s-1)! (2s-2)!}
 \sum_{j=1}^{\infty}\frac{j(2s+2j-3)!}{(2j+1)!}\left(
 \frac{\Delta v}{c}\right)^{2j-1}\right] ,
\end{eqnarray}
where we used denotation $S = 2\pi \omega_{oc}R_{I}/c = m_{oc}/M_{o}$.

   We can demonstrate that the power series in the result of integral
$I_{\vartheta}$ are convergent for $\Delta v/c < 1$. Condition
$\Delta v/c < 1$ sets the maximum possible acceleration of the particle
(i.e. maximum increment, $\Delta v$, acquired during the period $P_{o}$)
in the universe.

   Since $S = m_{oc}/M_{o} \ll 1$, usually, we can neglect higher powers
of $S$. As well, the common accelerations, $\Delta v$ within the period
$P_{o}$, are many orders of magnitude smaller than the speed of light,
$c$, therefore we can also neglect higher powers of $\Delta v/c$.
Integral $I_{\vartheta}$ for the all known situations in the universe
can be written in the simple form
\begin{equation}\label{iivartheta4}
I_{\vartheta} = \frac{2}{3}\frac{m_{oc}}{M_{o}}\left(
 \frac{\Delta v}{c}\right) + \frac{2}{5}\frac{m_{oc}}{M_{o}}
 \left( \frac{\Delta v}{c}\right)^{3} + ... \pm i\, \left[ \frac{2}
 {3}\frac{m_{oc}^{2}}{M_{o}^{2}}\left( \frac{\Delta v}{c}\right) +
 ... \right] .
\end{equation}
Since we observe only the phenomena manifestating themselves in the real
space, it is reasonable, also here, to consider only the real component
of $I_{\vartheta}$, where the second term containing $(\Delta v/c)^{3}$
can influence the size of inertia force in, e.g., an extreme deceleration
of a particle striking a high potential barrier. In the common physics,
it is enough to consider only the first term of the real component, with
which the inertia force is
\begin{equation}\label{ffi3}
F_{iner.} = -\frac{M_{o}K_{E}I_{t}}{3R_{I}^{2}} \frac{m_{oc}}{M_{o}}
 \frac{\Delta v}{c}.
\end{equation}
\indent \\

\normalsize
   REMARK. Notice that the imaginary part of integral $I_{\vartheta}$,
which corresponds to the hyperbolic cosine and, hence, the electric part
of the unified force, starts with the term contaning the second power of
$m_{oc}/M_{o}$. Actually, the integral of the first term is
\begin{equation}\label{ffifirst}
I_{\vartheta,i1} = \int_{0}^{\pi} (\pm i)\, \cos \vartheta \,
 \sin \vartheta \, d\vartheta = 0.
\end{equation}
{\bf This result explains why the inertia force is proportional
to the mass of bodies, but it is not proportional to the electric
charge} $[$even if the electric part $\pm \cosh (m_{oc}/M_{o})$ would be
re-scalled, in an alternative solution, to be real-valued$]$. We note
that typically $m_{oc}/M_{o} \approx 10^{-18}$, therefore the electric
,,inertia term'', $m_{oc}^{2}/M_{o}^{2}$, is about 18 orders of
magnitude smaller than the gravitational inertia term, $m_{oc}/M_{o}$.\\

\large
   The acceleration uses to be given in the form of $\Delta v/\Delta t$.
We said that the time interval, $\Delta t$, equals
$\Delta t = P_{o} = \hbar_{B}/(\pi M_{o}c^{2})$ in our case. So, we can
re-write the ratio $\Delta v/c$ as $(\Delta v/\Delta t)(\Delta t/c) =
[\hbar_{B}/(\pi M_{o}c^{3})](\Delta v/\Delta t)$. Using the latter, the
inertia force (its real component) can be given in the more familiar
form as
\begin{equation}\label{ffi4}
F_{iner.} = -\frac{\hbar_{B} M_{o}K_{E}}{3\pi}\frac{I_{t}}
 {R_{I}^{2}M_{o}^{2}c^{3}} m_{oc}\frac{\Delta v}{\Delta t}.
\end{equation}

   The inertia force of an object consisting of $N_{+}$ positively
charged particles having the mass $m_{+}$ and $N_{-}$ negatively
charged particles having the mass $m_{-}$ is simply the sum of
partial inertia forces of all constituent particles, i.e.
\begin{equation}\label{ffimacro}
F_{iner.} = -\frac{M_{o}K_{E}}{3\pi}\frac{\hbar_{B}I_{t}}
 {R_{I}^{2}M_{o}^{2}c^{3}} (N_{+}m_{+} + N_{-}m_{-})\frac{\Delta v}
 {\Delta t}.
\end{equation}

\section{Equation of motion}

\subsection{Equation of motion in traditional form}

   In two sections avove, the acting and inertia forces were derived in
forms different from those in the Coulomb and Newton laws. In this
subsection, we show that our result is consistent with the conventional
equations of motion containing the Coulomb and Newton laws despite
the difference.

   The inertia and acting forces should be equal in a description of
motion of a body. Let us now to consider the macroscopic test body being
in rest in the distance $r$ from another body, whereby the sizes of both
bodies are negligible in respect to the distance $r$. The equation of
motion of the test body, based on our derivations of acting and inertia
forces, is
\begin{eqnarray}\label{equmot1}
-\frac{M_{o}K_{E}}{3\pi}\frac{\hbar_{B}I_{t}}{R_{I}^{2}M_{o}^{2}c^{3}}
 (N_{+}m_{+} + N_{-}m_{-})\frac{\Delta v}{\Delta t} = \nonumber \\
 = \frac{M_{o}K_{E}<I_{\phi}>}{6r^{2}} \left[ -(n_{+} - n_{-})
 (N_{+} - N_{-}) + \right. \nonumber \\
 \left. + \frac{1}{M_{o}^{2}}(n_{+}m_{+} + n_{-}m_{-})
 (N_{+}m_{+} + N_{-}m_{-})\right] .
\end{eqnarray}
The forms $N_{+}m_{+} + N_{-}m_{-}$ and $n_{+}m_{+} + n_{-}m_{-}$ are,
in fact, the masses of the test object and acting object, which we
denote by $\tilde{m}_{T}$ and $\tilde{m}_{A}$, respectively. The forms
$(N_{+} - N_{-})M_{o}$ and $(n_{+} - n_{-})M_{o}$ can be given as
$(N_{+} - N_{-})q_{o}/\sqrt{4\pi \varepsilon_{o}\kappa}$ and
$(n_{+} - n_{-})q_{o}/\sqrt{4\pi \varepsilon_{o}\kappa}$. (We remind
that the elementary electromass
$M_{o} = q_{o}/\sqrt{4\pi \varepsilon_{o} \kappa}$.) Further, we denote
$\tilde{q}_{T} = (N_{+} - N_{-})q_{o}$ and
$\tilde{q}_{A} = (n_{+} - n_{-})q_{o}$, whereby $\tilde{q}_{T}$ and
$\tilde{q}_{A}$ are the electric charges of the test object and acting
object, respectively. The fine structure constant, $\alpha_{B}$, can be
given with the help of the gravitational constant, $\kappa$, elementary
electromass, $M_{o}$, Planck's constant divided by $2\pi$, and velocity
of light, $c$, as $\alpha_{B} = \kappa M_{o}^{2}/(\hbar_{B}c)$.

   After some handling, the equation of motion (\ref{equmot1}) can be
re-written into the form
\begin{equation}\label{equmot2}
\tilde{m}_{T}\frac{\Delta v}{\Delta t} = \frac{<I_{\phi}>}
 {8\pi I_{t}} \frac{1}{\alpha_{B}}\left[
 \frac{\tilde{q}_{T}\tilde{q}_{A}}{4\pi \varepsilon_{o}r^{2}} -
 \kappa \frac{\tilde{m}_{T}\tilde{m}_{A}}{r^{2}} \right] .
\end{equation}
If
\begin{equation}\label{condalpha}
\frac{<I_{\phi}>}{8\pi I_{t}} \frac{1}{\alpha_{B}} = 1,
\end{equation}
we obtain the unified equation of motion in the classical physics. In
more detail, if the charges $\tilde{q}_{T}$ and $\tilde{q}_{A}$ are not
zero, we can neglect the second term in the brackets and the equation
becomes
\begin{equation}\label{equmotel}
\tilde{m}_{T}\frac{\Delta v}{\Delta t} =
 \frac{\tilde{q}_{T}\tilde{q}_{A}}{4\pi \varepsilon_{o}r^{2}}.
\end{equation}
We can clearly see that it is the classical equation of motion for two
charged objects.

   If at least one charge is zero, then the first term in the brackets
of eq.(\ref{equmot2}) is zero and only the second term remains. So, the
equation, in this case, is
\begin{equation}\label{equmotgrav}
\tilde{m}_{T}\frac{\Delta v}{\Delta t} = 
 -\kappa \frac{\tilde{m}_{T}\tilde{m}_{A}}{r^{2}}.
\end{equation}
It is nothing else than the classical equation of motion for two
gravitationally interacting objects.

   The condition (\ref{condalpha}), i.e.
$<$$I_{\phi}$$>$$(1/\alpha_{B})/(8\pi I_{t}) = 1$, provides us with a
possibility to calculate the fine structure constant, if the perfect
model of the interaction is known. Specifically,
\begin{equation}\label{calcalpha}
\alpha_{B} = \frac{<I_{\phi}>}{8\pi I_{t}}.
\end{equation}
Or, the condition can be used to verify the correctness of the
interaction models in future studies. In our simple model of the impulse
absorption, with $<$$I_{\phi}$$> = (1 - 2/\pi )/\pi$ and
$I_{t} = 2/\pi$, we obtain
$\alpha_{Bcalc.} = (1 - 2/\pi )/(16\pi ) = 1/138.327511$. It is the
value about $1\%$ different from the actual experimental value of
$\alpha_{B} = 1/(137.035990 \pm 0.000006)$.

\subsection{Equation of motion in pure-geometrical form}

   The scientists who created the theory of relativity had an ambition
to work out the pure geometrical theory. It means that the equations
within this theory would be dimensionless. Our consistent derivation of
both acting and inertia forces allows us to achieve this goal, at least
in the case of the elementary equation of motion.

   In the absolutely autonomous description of a geometric structure,
the size of a given feature can be expressed, only, as a relative
multiple of the size of other feature. Or, we must choose a unit of
length, if we wish to speak about specific value of the size. If the
geometric structure is dynamical, changing shapes, sizes, and positions
of the features, we can again express the rate of a given change as a
relative multiple of the rate of other change. Or, we can do this with
establishing the definition of the unit of change. Therefore, two units
(relative or established to be ,,absolute'') are necessary, in
principle, in the description of pure geometrical, dynamical structure:
unit of the length and unit of the rate of its change.

   Let us to consider again the inertia force (\ref{ffi3}), in which we
generalize mass $m_{oc}$ to the mass of a macroscopic object, i.e. we
replace $m_{oc} \rightarrow N_{+}m_{+} + N_{-}m_{-}$. Putting this force
into the equality with the acting force in form (\ref{ffa}), we obtain
the primary form of the equation of motion:
\begin{eqnarray}\label{equmot3}
-\frac{M_{o}K_{E}}{3R_{I}^{2}}\frac{N_{+}m_{+} + N_{-}m_{-}}{M_{o}}
 \frac{\Delta v}{c} = \nonumber \\
 = \frac{M_{o}K_{E}<I_{\phi}>}{6r^{2}} \left[
 -(n_{+} - n_{-})(N_{+} - N_{-}) + \right. \nonumber \\
 \left. + \frac{1}{M_{o}^{2}}(n_{+}m_{+} + n_{-}m_{-})
 (N_{+}m_{+} + N_{-}m_{-})\right] ,
\end{eqnarray}
which can easily be simplified to
\begin{eqnarray}\label{equmot3b}
\left( N_{+}\frac{m_{+}}{M_{o}} + N_{-}\frac{m_{-}}{M_{o}}\right) 
 \frac{\Delta v}{c}
 = \frac{<I_{\phi}>}{2 I_{t}} \left( \frac{R_{I}}{r}\right)^{2} \left[
 (n_{+} - n_{-})(N_{+} - N_{-}) + \right. \nonumber \\
 \left. - \left( n_{+}\frac{m_{+}}{M_{o}} + n_{-}\frac{m_{-}}{M_{o}}
 \right) \left( N_{+}\frac{m_{+}}{M_{o}} + N_{-}\frac{m_{-}}{M_{o}}
 \right) \right] .
\end{eqnarray}
We derived this equation using the same initial formulas as in the
derivation of eq.(\ref{equmot2}). Only a difference in algebraic
operations was sufficient to eliminate the constants as gravitational
constant or permitivity of vacuum. Notice further that
eq.(\ref{equmot3b}) is dimensionless: only the dimensionless numbers
of particles, dimensionless integrals, and ratios of masses, velocities,
and distances figure there.

   In eq.(\ref{equmot3b}), the masses $m_{+}$ and $m_{-}$ can be
re-written with the help of the well-known de Broglie's relation as
$m_{+} = h_{B}\tilde{\nu}_{+}/c^{2}$ and
$m_{-} = h_{B}\tilde{\nu}_{-}/c^{2}$, where $\tilde{\nu}_{+}$ and
$\tilde{\nu}_{-}$ are the frequencies of the waves associated with
the particles of masses $m_{+}$ and $m_{-}$, respectively. As well, we
establish the frequency $\nu_{o}$ associated with the wave corresponding
to the elementary electromass, $M_{o}$, whereby
$M_{o} = h_{B}\nu_{o}/c^{2}$. Or, the frequencies can be replaced with
the corresponding wavelengths when we use relations
$\tilde{\nu}_{+} = c/\tilde{\lambda}_{+}$,
$\tilde{\nu}_{-} = c/\tilde{\lambda}_{-}$, and
$\nu_{o} = c/\Lambda_{C}$. The ratio of integrals
$<$$I_{\phi}$$>$$/I_{t}$ can be given with the help of relation
(\ref{condalpha}) as $<$$I_{\phi}$$>$$/I_{t} = 8\pi \alpha_{B}$.
Using the above mentioned relations and after few algebraic operations,
the equation (\ref{equmot3b}) can be re-written into the form
\begin{eqnarray}\label{equmot5}
\left( N_{+}\frac{\Lambda_{C}}{\tilde{\lambda}_{+}} + N_{-}
 \frac{\Lambda_{C}}{\tilde{\lambda}_{-}}\right) \frac{\Delta v}{c} =
 4\pi \alpha_{B} \left( \frac{R_{I}}{r}\right) ^{2}
 \left[ (n_{+} - n_{-})(N_{+} - N_{-}) - \right. \nonumber \\
 \left. - \left( n_{+}\frac{\Lambda_{C}}{\tilde{\lambda}_{+}} + n_{-}
 \frac{\Lambda_{C}}{\tilde{\lambda}_{-}}\right) \left( N_{+}
 \frac{\Lambda_{C}}{\tilde{\lambda}_{+}} + N_{-}\frac{\Lambda_{C}}
 {\tilde{\lambda}_{-}}\right) \right] .
\end{eqnarray}

   The last equation contains only the dimensionless ratios $r/R_{I}$,
$\tilde{\lambda}_{+}/\Lambda_{C}$, $\tilde{\lambda}_{-}/\Lambda_{C}$,
and $\Delta v/c$. The ratio $r/R_{I}$ gives the distance $r$ as the
multiple of the interaction radius, $R_{I}$. $[$The interaction radius
may not necessarily be the unit of length. In principle, we can
establish the unit of length $R_{u}$ and replace $r/R_{I}$ with
$(r/R_{u})(R_{u}/R_{I}).]$ The ratio $\tilde{\lambda}_{+}/\Lambda_{C}$
($\tilde{\lambda}_{-}/\Lambda_{C}$) gives how many times is the
wavelength $\tilde{\lambda}_{+}$ ($\tilde{\lambda}_{-}$) of the wave
associated with the particle of mass $m_{+}$ ($m_{-}$) longer than the
wavelength associated to elementary electromass $M_{o}$. The ratio
$\Delta v/c$ says how large is the rate of the change of test-particle
position in the inertial frame, in which the particle is situated in
the beginning of a detection interval, in comparison with the change of
the position of a photon within the detection interval.

   We demonstrated that our equation of motion, with mutually consistent
acting and inertia forces, which can be written in the traditional form
with the Coulomb and Newton laws, can also be re-written as the
dimensionless equation, without physical constants as the gravitational
constant, permitivity of vacuum, or charge. In this context,
{\bf the physical constants appear to be the transformation
constants} between the natural physical quantities/units and
historical, artificial quantities/units established by man in the past,
when the knowledge was not advanced enough for doing the descripton of
the physical phenomena in the natural way.

   The constant $4\pi \alpha_{B}$ in Eq.(\ref{equmot5}) is the
mathematical constant, i.e. the absolute constant from the point of view
of physics. It must therefore be the same in whatever time and whatever
universe. In the context of this constancy, let us to discuss the strong
equivalence principle (SEP) in relation to the presented concept of the
unique interaction.

   Considering the complete description of the whole interaction and
realizing that the amplitude of this unique interaction is proportional
to cosine and sine (hyperbolic cosine and hyperbolic sine) functions,
the SEP is clearly violated, in our concept. This had to occur, since
in our re-gauging of the integration constant $C_{2}$ in the
Schwarzschild solution of Einstein field equations, in Sect.~2, we
explicitly assumed, when replacing the potential with potential energy,
that the metrique of spacetime is modulated not only by the central,
but the test particle as well.

   Our replacement of the potential with potential energy, causing the
violation of the SEP, is in agreement with the quantum-physics
description of microcosm (with the Schr\"{o}dinger, Klein-Gordon, or
Dirac's equations in particular), where the potential energy, but never
the potential, figures. The description is consistent with the observed
reality, therefore it is clear that the SEP cannot be valid in the
domain of quantum phenomena. And, because every actually unified theory
must include also these phenomena, it cannot satisfy the SEP.

   We however know that experiments have proved, till now, the validity
of the SEP in the case of the gravitational, macroscopic interaction.
The latter corresponds with the objects consisting of exactly the same
numbers of positively and negatively charged elementary sources of
waving, in our concept. Since the amplitude of the interaction between
such the objects is described by hyperbolic sine, not by only a single
constant, its behavior is not exactly proportional to the $1/r^{2}$ law.
Thus, the SEP must be, in principle, violated in our concept also in
this case. Despite this fact, one can state, within the presented
hypothesis, that the SEP occurs to be a very good approximation of true
description of reality, when applied only to the gravitational part of
the unique force.

   Two following circumstances justify the correctness of the
approximation. At first, the constant of the proportionality in the
force law (see Eq.(\ref{equmot5})), $4\pi \alpha_{B}R_{I}^{2}$, which is
the equivalent of the gravitational constant in our new concept, is true
constant. (We know that this constancy is the consequence of the SEP.)
In our concept, $\alpha_{B}$ is the mathematical (absolute) constant and
we have no reason to suppose any change of interaction radius,
$R_{I}$ (although, the concept itself would remain valid even if $R_{I}$
was a function of time).

   At second, the deviation of gravity from the $1/r^{2}$ behavior, and
thus from the SEP requirements, is extremely small. The amplitude of
the gravitational part of the unique interaction in the domena of
macroscopic phenomena is proportional to $\sinh (|kr|)$ (i.e. to the
gravitational part of absorption amplitude at the distance $r$ between
the interacting particles). This function can be approximated by the
first term of expansion to the power series, $|kr|$, since this argument
is much smaller than unity. For $r \ll R_{S}$, it can be given as
$|kr| = (2\pi \omega_{o}r/c)/\sqrt{1 - r/R_{S}} =
(2\pi \omega_{o}r/c)[1 - r/(2R_{S}) + ...]$, where $r/(2R_{S})$ is the
term causing the difference of gravity from the $1/r^{2}$ law. For the
macroscopic bodies interacting only gravitationally, $R_{S}$ is very
large and, consequently, the ratio $r/(2R_{S})$ is such a small value
that cannot be detected in any current or near-future experiment. For
example, $R_{S}$ is of the order of $\sim$$10^{30}\,$m for the system
consisting of the Sun and a spacecraft with the mass of $10\,$kg, which
can carry a measurement apparatus for general-relativity experiments
and which modulates the surrounding spacetime together with the Sun. We
consider the Sun as the central body, since its gravitational potential
energy is the highest potential energy in the cosmic space in the
Earth's region as well as on the Earth's surface. The above calculated
potential energy corresponds to a location of the spacecraft far from
the Earth, where the contribution of this planet to the spacetime
modulation can be neglected (we consider this situation to be
conservative in calculating $R_{S}$). If the distance of the spacecraft
from the Sun is $\sim$$1$ astronomical unit, i.e. of order of
$\sim$$10^{11}\,$m, the term $r/(2R_{S})$ for the Sun-spacecraft system
is $\sim$$10^{-19}$. So, the deviation from the SEP occurs at the 19-th
decimal digit, in this case. Clearly, such the deviation cannot be
measured.

\section{Summary}

   The unification of the fundamental interacton in the universe seems
to be possible, if we assume that the universe consists, in its deepest
elementary level of existence, of the elementary sources that generate
the waving, each its own, spreading in the surronding space.

   The basic characteristics of the elementary sources of waving can be
associated with the matematical objects, which are well-known complex
numbers. The electric charge is associated with the imaginary-valued
component of the complex number and mass with its real-valued component.
The complex number and its conjugate, i.e. the complex number with
opposite sign of its imaginary-valued component, correspond to the
electric charges of two polarities.

   In the description of the elementary source of waving, the cosine and
sine or hyperbolic cosine and hyperbolic sine, which are the parts of
common exponential function, are the main mathematical feature. It
appears to be natural to assign the cosine (hyperbolic cosine) to the
amplitude of electric existence of wave source and the sine (hyperbolic
sine) to the amplitude of gravitational existence of the wave source.
The cosine and sine describe the source in microcosm and hyperbolic
cosine and hyperbolic sine in macrocosm. In the macroscopic environment
we live in, the argument of these hyperbolic functions is much lower
than unity and this fact implies a much lower amplitude of
sine-gravitational force in comparison with the amplitude of the
cosine-electric force.

   After a developing of the functions into the power series,
$\cosh (x) = 1 + x^{2}/2! + ...$, $\sinh (x) = x + x^{3}/3! + ...$,
hyperbolic cosine always starts with unity (which does not depend on
the argument). Or, it can be put to equal to unity, because higher terms
can be neglected due to the value of the argument being negligible with
respect to unity. This fact allows us to explain several qualitative
properties of elementary particles. The first is that the magnitude of
elementary electric charge has been measured the same for all known
electrically charged elementary particles. On contrary, the mass
described by $\sinh (x) \doteq x + ...$, which thus depends on the
variable argument, $x$, is actually correspondingly different for
various elementary particles.

   Schematically, a positively charged particle can be related to the
sum $+i A \cosh(x) + B \sinh (x) = iA + Bx +...$ and a negatively
charged particle to $-i A \cosh (x) + B \sinh (x) = -iA + Bx -...$.
Symbols $A$ and $B$ stand for the constants for which we showed the
validity of sharp inequality $|A| \gg |Bx|$ for every value of $x$ in
the domain of macroscopic phenomena. Adding the same numbers ($N$) of
the sums of both positive and negative charges
$[N(+iA + Bx + ...) + N(-iA + Bx -...)]$, which correspond to the same
numbers of both positively and negatively charged particles, diminishes
the absolute members, i.e. $iA$ and $-iA$. Therefore, the electrically
neutral object can exist. The second members of the developments of
whatever-argument hyperbolic sines, which correspond to mass, are,
however, only additive, lifting the final sum (to $2NBx$), therefore
the mass cannot be ,,neutralized''. Gravity is, consequently, always
present at the material bodies.

   The fact that electric charge is proportional to $\pm iA$, i.e. the
constant (after development of hyperbolic cosine into the power series
and neglection of higher terms), also implies that a relativistically
moving charge does not depend on velocity or a curvature of space-time
like a mass. There exists the well-known formula
$m = m_{o}/\sqrt{1 - v^{2}/c^{2}}$ for the mass $m$ of a moving object
($m_{o}$ is its rest mass), but no any analogous formula for the
electric charge. In principle, the dependence of charge on the
velocity/curvature exists. However, it becomes apparent for values of
factor $1/\sqrt{1 - v^{2}/c^{2}}$ not comparable or larger than unity
(like in the case of mass), but comparable or larger than ratio
$M_{o}/m_{o}$ (which is about $\sim$$10^{18}$ for a proton and
$\sim$$10^{21}$ for an electron).

   Regardless whether the ,,mass'' of the wave source is defined as
positive or negative quantity (the sign of mass is a matter of
convention; we could, in principle, establish the mass as a
negatively-valued quantity), the product of multiplication of such two
values is always positive:\\
$(+m).(+m) = +m^{2}$;\\
$(-m).(-m) = +m^{2}$.\\
Therefore there can be only a single orientation of the gravity action
$[$either attractive or repulsive; in means of pure theory, we are not
able to determine the appropriate integration constant ($K_{E}$ in our
denotation) and, thus, predict whether the gravity is attractive or
repulsive; this must be done in experiment$]$. The product of
multiplication of two imaginary units, $i$, of the same sign $[$opposite
signs$]$, which corresponds with the orientation of the elecric force
between the charges of the same $[$opposite$]$ polarity, is\\
$(+i).(+i) = -1$ or $(-i).(-i) = -1$,\\
$[(+i).(-i) = +1$ or $(-i).(+i) = +1]$.\\
Since the integration constant is identical, the orientation of force
can differ only due to a difference of the product of this
multiplication. Thus, we can predict the orientation of the electric
force between the given pair of charges with respect to the orientation
of gravitational force. Actually, the orientation between the charges
of opposite polarity is the same as the orientation of gravity.

   We pointed out the very probable mechanism leading to the occurrence
of inertia force: the increase of the frequency of the wave associated
with a given particle in the direction of the particle acceleration and
decrease of the frequency in the opposite direction, which cause an
inequality of the impulse delivered to the particle by its own wave.
The first, absolute term of the electric (hyperbolic cosine) part, i.e.
$\pm iA$, does not depend on the frequency. Consequently, the inertia
force does not depend on an electric charge (or depends only on its
higher, negligible-in-common-world terms). If the increment of the
velocity of particle during the period of one oscillation of wave
associated with the elementary electromass is negligible in comparison
with the velocity of light (which is true in our, common world), the
inertia force is linearly proportional to the acceleration. In an
opposite case, at extreme accelerations, the force increases more
steeply and approaches infinity, so far, when the above mentioned
velocity increment approaches the speed of light, $c$. In reality,
there is not only the speed limit, velocity of light, but the limit of
acceleration as well.

   Full formula giving the inertia force also contains the higher than
linear powers of $m/M_{o}$ to which the force is proportional. In our
common environments, $m/M_{o} \ll 1$, therefore these higher powers can
be neglected. In an ultra relativisic, extremely curved space-time
(e.g. inside an object which is just going to become a black hole),
ratio $m/M_{o}$ can however approach unity or to be much larger than
unity, therefore the inertia force is predicted to very steeply
increase even at small accelerations, there.

   For two objects charged with the charges of opposite polarity, we
found the region, where the force does not change monotonously with
the distance, but oscillates, i.e. it changes its orientation. In this
region, there are ,,levels'' of zero force in which the particles can
be in a static stable configuration. In other words, our description of
the unified force predicts the existence of stable, static, bound
structures, which are actually observed as atoms or molecules. In more
detail, the zero-force levels are predicted by the solution of the
quadratic equation with two roots. These two solutions imply two zones
of stable, zero-force levels, the first of which corresponds to the
atom shell (and, likely, molecular bounds after a wider generalization)
and the second zone corresponds to the stable configurations at the
atom nucleus.

   The border between the monotonous and oscillating behavior of the
unique force was revealed as the consequence of our new definition
(re-gauging) of the constants in the Schwarzschild's solution of
the Einstein field equations. This new gauging of the constants, which
is the essential assumption of our thery, could be verified in
experiments since it should be possible to construct a macroscopic-size
atom. Namely, if we construct a point-like, positively charged object,
with charge $q_{C}$ and ,,spray'' few electrons of the net charge
$q_{T}$ in its vicinity, then the size of generalized Schwarzschild
radius, $R_{S}$, can be controlled tuning the charge $q_{C}$, whereby
\begin{equation}\label{qcc}
R_{S} = \frac{q_{T}}{4\pi \varepsilon_{o}m_{e}c^{2}}{q_{C}}.
\end{equation}
For example, the radius can be 20 centimeters, if
$q_{T} \sim -10^{1}\, q_{o}$ and $q_{C}$ is set to be
$\sim$$1.1\times 10^{-6}\,$C. The electrons ,,sprayed'' around the
central object should be trapped in the zero-force distances at $R_{S}$
and such their assembly should behave in a way that could help us to
visualize the oscillating force.

   Such the macroscopic atom would provide us also with a unique
possibility to study, on the macroscopic-dimension scale, the phenomena,
which we call quantum phenomena. One can guess that this is the
principle of the occurrence of the weird phenomenon, which is known as
,,ball lightning''.\\
\vspace{5 mm}

   {\bf Acknowledgements.} I thank the organizers of the workshop
,,GOING BEYOND METRIC: Black Holes, Non-Locality and Cognition'',
Dr. Metod Saniga in particular, for providing me with the opportunity
to present the result of my work on the unified-interaction theory
within this workshop.\\

\end{document}